\documentclass[
twocolumn, 
prb, 
aps, 
amsmath,
amssymb,
superscriptaddress, 
showkeys
]{revtex4-2}

\usepackage{graphicx}
\usepackage{bm}
\usepackage[dvipsnames]{xcolor}
\usepackage{color}
\usepackage{siunitx}
\usepackage{natmove}
\usepackage{float}
\usepackage{mathtools}
\usepackage{hyperref} 
\usepackage[capitalise]{cleveref} 
\usepackage[normalem]{ulem} 
\usepackage{booktabs}

\usepackage{natbib}
\usepackage{natmove}
\usepackage{cancel}

\newcommand{\beq}{\begin{equation}}
\newcommand{\eeq}{\end{equation}}
\newcommand{\bea}{\begin{eqnarray}}
\newcommand{\eea}{\end{eqnarray}}

\newcommand{\ed}[1]{{\color{black}{#1}}}

\global\csname @topnum\endcsname 0
\global\csname @botnum\endcsname 0

\begin{document}

\title{Ultra-fast interpretable machine-learning potentials}
\author{Stephen R. Xie}
\affiliation{Department of Materials Science and Engineering, University of Florida}
\affiliation{Quantum Theory Project, University of Florida}
\author{Matthias Rupp}
\affiliation{Department of Computer and Information Science, University of Konstanz, Germany}
\affiliation{Present address: Luxembourg Institute of Science and Technology (LIST), Luxembourg}
\author{Richard G. Hennig}
\affiliation{Department of Materials Science and Engineering, University of Florida}
\affiliation{Quantum Theory Project, University of Florida}
\date{incomplete draft version of \today}

\begin{abstract}\noindent
All-atom dynamics simulations are an indispensable quantitative tool in physics, chemistry, and materials science, but large systems and long simulation times remain challenging due to the trade-off between computational efficiency and predictive accuracy.
To address this challenge, we combine effective two- and three-body potentials in a cubic B-spline basis with regularized linear regression to obtain machine-learning potentials that are physically interpretable, sufficiently accurate for applications, as fast as the fastest traditional empirical potentials, and two to four orders of magnitude faster than state-of-the-art machine-learning potentials.
For data from empirical potentials, we demonstrate exact retrieval of the potential.
For data from density functional theory, the predicted energies, forces, and derived properties, including phonon spectra, elastic constants, and melting points, closely match those of the reference method.
The introduced potentials might contribute towards accurate all-atom dynamics simulations of large atomistic systems over long time scales.
\end{abstract}


\maketitle{}


\section{Introduction}

All-atom dynamics simulations enable the quantitative study of atomistic systems and their interactions in physics, chemistry, materials science, pharmaceutical sciences, and related areas. The simulations' capabilities and limits depend on the potential used to calculate the forces acting on the atoms, with an inherent correlation between the accuracy of the underlying physical model and the required computational effort: On the one hand, electronic structure methods tend to be accurate, slow, applicable to many systems, and require little human parametrization effort. On the other hand, traditional empirical potentials are fast but limited in accuracy and applicability, with often high parametrization effort.

\begin{figure}[H]
    \centering
    \includegraphics[width=\columnwidth]{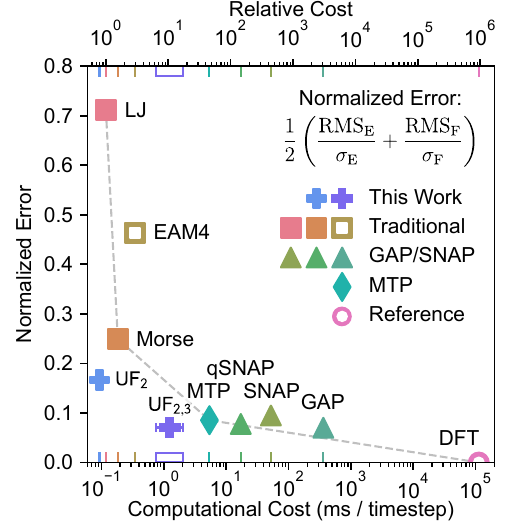}
    \caption{%
        \emph{Trade-off between prediction error and computational cost} of evaluating machine-learning potentials.
        Prediction errors are relative to the underlying electronic-structure reference method (disk).
        Ultra-fast potentials (this work, stars) with two-body (UF$_2$) and three-body (UF$_{2,3}$) terms are as fast as traditional empirical potentials (squares) but close in error to state-of-the-art machine-learning potentials (triangles, diamond).
        All potentials except EAM4 were refitted to the same tungsten data set. \ed{Computational costs were benchmarked with a 128-atom bcc-tungsten supercell.}
        Trade-offs between accuracy and cost also arise from the choices of hyperparameters for each potential.
        See Sections~\ref{sec:background} and \ref{sec:methods} for abbreviations and details.
    }
    \label{fig:pareto}
\end{figure}

Machine-learning potentials (MLPs)~\cite{Deringer2019, Langer2022, Miksch2021, Friederich2021} are flexible functions fitted to reference energy and force data from, e.g., electronic structure methods. Their computational advantage does not primarily arise from simplified physical models but from avoiding redundant calculations through interpolation. Hence, they can stay close to the accuracy of the reference method while being orders of magnitude faster (see \cref{fig:pareto}), with little human parametrization effort, but are often hard to interpret. However, current accurate MLPs are still orders of magnitude slower than traditional empirical potentials, limiting their use for dynamics simulations of large atomistic systems over long time scales.

In this work, we develop an interpretable linear MLP based on effective two- and three-body potentials using a flexible cubic B-spline basis. \Cref{fig:pareto} demonstrates how this ultra-fast potential (UF) is close in error to state-of-the-art MLPs while being as fast as the fastest traditional empirical potentials, such as the Morse and Lennard-Jones potentials.

\section{Background}
\label{sec:background}

\emph{The many-body expansion} \cite{Drautz2004} of an atomistic system's potential energy
\begin{multline}
    \label{eq:manybodyexpansion}
    E = E_0 + \frac{1}{1!} \sum_{i} V_1({\bf R}_i,\sigma_i) + \frac{1}{2!} \sum_{i,j} V_2({\bf R}_i, \sigma_i, {\bf R}_j, \sigma_j) \\
        + \frac{1}{3!} \sum_{i,j,k} V_3({\bf R}_i,\sigma_i,{\bf R}_j,\sigma_j,{\bf R}_k,\sigma_k) + \dots
\end{multline}
is a sum of $N$-body potentials~$V_N$ that depend on atom positions~${\bf R}_i$ and element species~$\sigma_i$, where the prefactors account for double-counting. Assuming transferability of the $V_N$ across configurations provides the basis for empirical and machine-learning potentials.

In this study, we omit the species dependency as well as the reference energies $E_0$ and $\frac{1}{1!} \sum_i V_1({\bf R}_i,\sigma_i)$ and subsume the factorial prefactors into the potentials~$V$ for simplicity. The extension to multi-component systems is straightforward and implemented in the accompanying program code.

\emph{Traditional empirical potentials} often have rigid functional forms with a small number of tunable parameters. These are optimized to reproduce experimental quantities, such as lattice parameters and elastic coefficients, as well as calculated quantities from first principles, such as energies of crystal structures, defects, and surfaces~\cite{rapaport2004,Martinez2013, Ragasa2019}. Typically this requires global optimization, e.g., via simulated annealing and substantial human effort.

\emph{Pair potentials} truncate \cref{eq:manybodyexpansion} after two-body terms,
\begin{equation}
    \label{eq:pairpotential}
    E = \frac{1}{2} \sum_{i,j} V_2(r_{ij}),
\end{equation}
where~$r_{ij}$ is the distance between atoms~$i$ and~$j$. These potentials are limited to systems where higher-order terms such as angular and dihedral interactions are negligible. The functional forms of the Lennard-Jones (LJ)~\cite{Jones1924} potential
\begin{equation}
  \label{eq:lj}
  V^{\text{LJ}}(r_{ij}) = 4 \epsilon \left [ \left ( \frac{\sigma}{r_{ij}} \right )^{12} - \left ( \frac{\sigma}{r_{ij}} \right )^6 \right ]
\end{equation}
and the Morse potential~\cite{Morse1929}
\begin{equation}
  \label{eq:morse}
  V^{\text{Morse}}(r_{ij}) = D_0 \left (e^{-2 a(r_{ij}-r_c)} - 2 e^{-a(r_{ij}-r_c)} \right )
\end{equation}
were originally developed for their numerical efficiency, where $\epsilon, \sigma$, and $D_0, a, r_c$ are model parameters.

\emph{Many-body potentials} extend the pair formalism by including additional many-body interactions, either in the form of many-body functions as in \cref{eq:manybodyexpansion}, or via environment-dependent functionals, such as in the embedded atom method (EAM)~\cite{Daw1984},
\begin{equation}
  \label{eq:eam}
  E^{\text{EAM}} = \frac{1}{2} \sum_{i, j} V(r_{ij}) + \sum_i F \Bigl( \sum_{j\ne i}\rho(r_{ij}) \Bigr) .
\end{equation}
Here, the embedding energy $F$ is a non-linear function of the electron density $\rho$, which is approximated by a pairwise sum. 

While traditional empirical potentials have seen success in applications across decades of research, their rigid functional forms limit their accuracy. More recently, MLPs with flexible functional forms and built-in physics domain knowledge in the form of engineered features or deep neural network architectures have emerged as an alternative~\ed{\cite{Langer2022,mgcc2021q,dbcc2021q,hl2021q,b2021q,uctm2021q}}
State-of-the-art MLPs can simulate the dynamics of large (``high-dimensional'') atomistic systems with an accuracy close to the underlying electronic-structure reference method but orders of magnitude faster. \ed{\cite{Parsaeifard2020,Zuo2020}}
However, they are still orders of magnitude slower than fast traditional empirical potentials, limiting their application in system size and simulation length.

Recent efforts to improve speed and accuracy of MLPs include using linear regression models, which can be faster to train and evaluate than non-linear models~\cite{Lysogorskiy2021, Kovacs2021}.
The spectral neighbor analysis potential (SNAP)~\cite{Thompson2015} and its quadratic variant (qSNAP)~\cite{Wood2018} are linear models based on the bispectrum representation~\cite{Bartok2009}.
Moment tensor potentials (MTP)~\cite{Shapeev2016}, 
atomic cluster expansion potentials~\cite{Drautz2019}, 
atomic permutationally-invariant polynomials (aPIP) potentials~\cite{VanderOord2019},
\ed{and Chebyshev interaction model for efficient simulation (ChIMES) potentials \cite{lindsey2017}}
are linear models based on polynomial basis sets.

A complementary approach to improve speed is to use basis functions that are fast to evaluate. In the context of MLPs, non-linear kernel-based MLPs have been trained and subsequently projected onto a spline basis, yielding a linear model~\cite{Vandermause2020}. Similar to this work, the general two- and three-body potential (GTTP) approach employs a quadratic spline basis set, exceeding MTPs in speed when trained on the same data~\cite{pozdnyakov2020}. Polynomial symmetry functions (PSF) improve over Behler-Parrinello symmetry functions \cite{Behler2007} in speed and accuracy by introducing compact support~\cite{Bircher2021}. Recently, new methods for fitting spline-based modified EAM potentials were benchmarked against MLPs, demonstrating comparable accuracy despite the lower complexity of their functional forms~\cite{Vita2021}.

Motivated by these observations, we developed an ultra-fast (UF) MLP that combines the speed of the fastest traditional empirical potentials with an accuracy close to state-of-the-art MLPs by employing regularized linear regression with spline basis functions with compact support to learn effective two- and three-body interactions. \Cref{fig:pareto} showcases the relation between prediction errors and computational costs for three traditional empirical potentials (LJ, Morse, EAM) and several MLPs benchmarked on a dataset of elemental tungsten~\cite{Szlachta2014}. While the traditional empirical potentials are fast but limited by accuracy, the MLPs are accurate but limited by speed.
UF~MLPs improve on the Pareto frontier of predictive accuracy and computational costs.
They are available as an open-source Python implementation (UF$^3$, Ultra-Fast Force Fields)~\cite{xie2021uf} with interfaces to the VASP~\cite{kresse1996} electronic-structure code
and the LAMMPS~\cite{Plimpton1995,tabbbcvkmnsetal2022} molecular dynamics code.


\section{Results and Discussion}

\begin{figure}[tb]
    \includegraphics[width=\columnwidth]{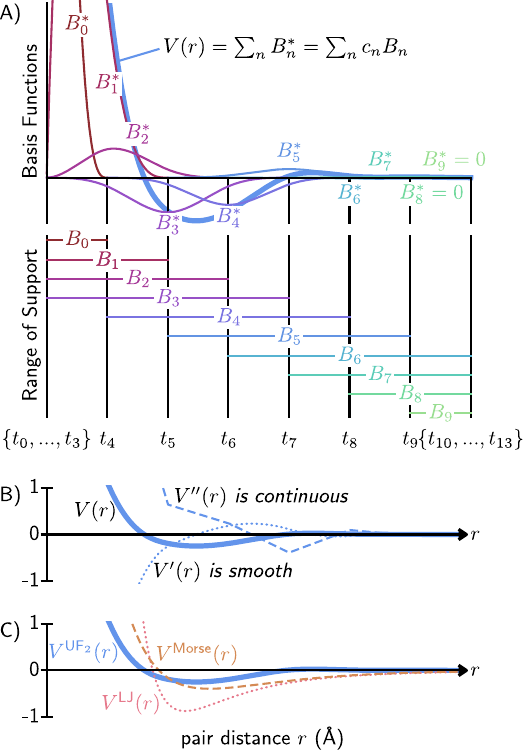}
    \caption{
        \emph{B-spline basis set for pair potentials.}
        A) Ten weighted B-splines after fitting ($B^*_n=c_n B_{n}(r)$, solid curves) and their ranges of support. Their sum is the fitted pair potential ($V(r)$, blue curve). In this example, knots $t_n$ are selected with uniform spacing and illustrated as vertical lines.
        B) Using cubic B-spline basis functions, the pair potential $V(r)$ has a smooth and continuous first derivative $V^\prime(r)$ (dotted line), which is essential for reproducing accurate forces. Its second derivative $V^{\prime\prime}(r)$ (dashed line) is continuous, which is essential for reproducing stresses and phonon frequencies.
        C) The optimized UF$_2$ potential exhibits more inflection points than the optimized LJ (dotted line) and Morse (dashed line) potentials, highlighting its increased flexibility.
        }
        \label{fig:workflo}
    
\end{figure}

\begin{figure*}[tb]
    \includegraphics[width=\textwidth]{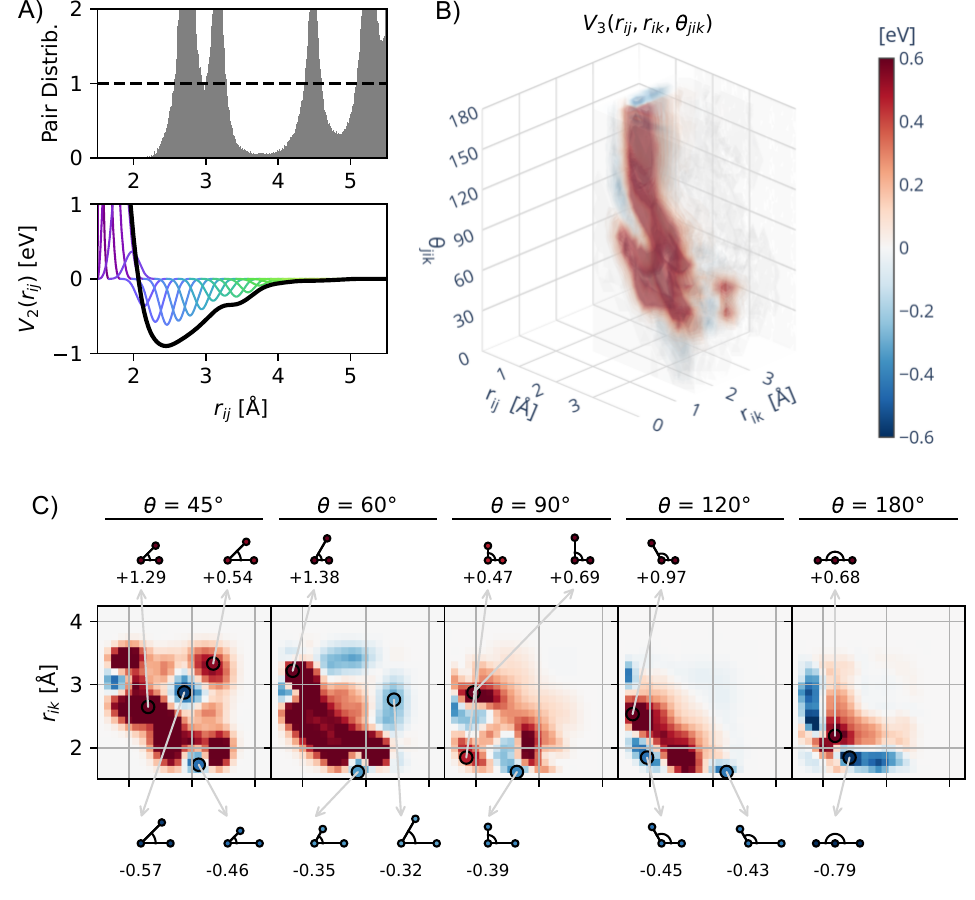}
    \caption{
        \ed{
        \emph{Visualization of two-body and three-body terms of a tungsten UF potential.} 
        a) Distribution of pair interactions in tungsten training data and the learned two-body component of the fitted potential.
        b) The learned three-body component $V_3(r_{ij}, r_{ik}, r_{jk})$ corresponds to the contribution to the total energy of a central atom $i$ interacting with two neighbors $j$ and $k$. It is fit simultaneously with the two-body potential. Here, $\theta_{jik}$ is substituted for $r_{jk}$ using the law of cosines for ease of visualization.
        c) Volume slices of $V_3(r_{ij}, r_{ik}, \theta_{jik})$ reveal favorable (blue) and unfavorable (red) three-body interactions.
        }
        }
        \label{fig:interpret}
\end{figure*}

The central idea of UF MLPs is to learn an effective low-order many-body expansion of the potential energy surface, using basis functions that are efficient to evaluate. For this, we truncate the many-body expansion of \cref{eq:manybodyexpansion} at two- or three-body terms and express each term as a function of pairwise distances (one distance for the two-body and three distances for the three-body term). This approach is general and can be extended to higher-order terms.
To minimize the computational cost of predictions, we represent $N$-body terms in a set of basis functions with compact support and sufficiently many derivatives to describe energies, forces, and vibrational modes, i.e., cubic splines.

\subsection{Expansion in B-splines}

\emph{Splines} are piecewise polynomial functions with locally simple forms, joined together at knot positions~\cite{boor2001}. They are globally flexible and smooth but do not suffer from some of the oscillatory problems of polynomial interpolators (e.g., Runge's phenomenon \cite{Runge1901}).

Spline interpolation is well-established in empirical potential development \cite{Wolff1999,Wen2015}. The LAMMPS package \cite{Plimpton1995}, a leading framework for molecular dynamics simulations, implements cubic spline interpolation for pair potentials and selected many-body potentials, including EAM. The primary motivation for splines is the desire for computational efficiency and the opportunity to improve accuracy systematically~\cite{Hennig2008}. Their compact support and simple form make spline-based potentials the fastest choice to evaluate, and adding more knots increases their resolution.

\emph{B-splines} constitute a basis for splines of arbitrary order. They are recursively defined as \cite{boor2001}
\begin{align}
B_{n, 1}(r) & = 
\begin{cases}
1, & t_n \leq r < t_{n+1} \nonumber \\
0, & \text{otherwise}
\end{cases}
\\
\label{eq:b-splines}
B_{n,d+1}(r) & = \frac{r-t_n}{t_{n+d}-t_n}B_{n,d}(r) \nonumber \\
& \quad + \frac{t_{n+d+1} - r}{t_{n+d+1} - t_{n+1}}B_{n+1, d}(r),
\end{align}
where $t_n$ is the $n$-th knot position, and $d$ is the degree of the polynomial. The position and number of knots, a non-decreasing sequence of support points that uniquely determine the basis set, may be fixed or treated as free parameters. B-Splines are well suited for interpolation due to their intrinsic smoothness and differentiability. Their derivatives are also defined recursively:
\begin{equation}
B^\prime_{n,d+1}(r) = d \bigg(\frac{B_{n,d}(r)}{t_{n+d}-t_n} - \frac{B_{n+1, d}(r)}{t_{n+d+1} - t_{n+1}}\bigg)
\end{equation}

The UF potential describes the energy~$E$ of an atomistic system via two- and three-body interactions:
\begin{equation}
    \label{eq:potential}
    E = \sum_{i,j} V_2( r_{ij}) + \sum_{i,j,k} V_3(r_{ij},r_{ik},r_{jk}) .
\end{equation}
For finite systems such as molecules or clusters, indices $i,j,k$ run over all atoms.
For infinite systems modeled via periodic boundary conditions, $i$ runs over the atoms in the simulation cell, and $j,k$ run over all neighboring atoms, including those in adjacent copies of the simulation cell.
While these are infinitely many, the sums are truncated in practice by assuming locality, that is, finite support of $V_2$ and $V_3$.

Modeling $N$-body interactions requires $n \choose 2$-dimensional tensor product splines.
The UF potential therefore expresses $V_2$ and $V_3$ as linear combinations of cubic B-splines, $B_{n} = B_{n,3+1}$, and tensor product splines:
\begin{align}
    V_2(r_{ij}) = & \sum ^{K}_{n=0} c_n B_{n}(r_{ij}) \nonumber \\
    V_3(r_{ij}, r_{ik}, r_{jk}) = & \nonumber \\ \sum ^{K_l}_{l=0} \sum ^{K_m}_{m=0} \sum ^{K_n}_{n=0} & c_{lmn} B_{l}(r_{ij}) B_{m}(r_{ik}) B_{n}(r_{jk}),
\end{align}
where $K$, $K_l$, $K_m$, and $K_n$ denote the number of basis functions per spline or tensor spline dimension, and $c_n$ and $c_{lmn}$ are corresponding coefficients.

The B-spline basis set spans a finite domain and is bounded by the end knots $[t_0, t_{K}]$. At the upper limit $t_{K}$, the potential smoothly goes to zero, and near the lower limit $t_0$, it monotonically increases with shorter distances to prevent atoms from getting unphysically close. \Cref{fig:workflo}A illustrates the compact support of the cubic B-spline basis functions. By construction, each basis function is nonzero across four adjacent intervals. Therefore, evaluating the two- and three-body potentials involves evaluating at most 4 and $4^3=64$ basis functions for any pair or triplet of distances, respectively, giving rise to the aforementioned computational efficiency.

The force ${\bf F}_a$ acting on atom $a$ is given as the negative gradient $-\nabla_{{\bf R}_a} E$ of the energy $E$  with respect to the atom’s Cartesian coordinate ${\bf R}_{a}$ and obtained analytically from the derivatives of the two- and three-body potentials,

\begin{align}
    \frac{\partial V_2(r_{ij}) }{\partial R_{a,\ell}} & = \sum ^{K_n}_{n=0} c_{n} B^\prime_{n}(r_{ij})\frac{\partial r_{ij}}{\partial R_{a,\ell}} \nonumber \\
    \frac{\partial V_3(r_{ij}, r_{ik}, r_{jk}) }{\partial R_{a,\ell}} & = \sum ^{K_l}_{l=0} \sum ^{K_m}_{m=0} \sum ^{K_n}_{n=0} c_{lmn} \bigg( \nonumber \\
    B^\prime_{l}(r_{ij})& B_{m}(r_{ik}) B_{n}(r_{jk})\frac{\partial r_{ij}}{\partial R_{a,\ell}} + \nonumber \\
    B_{l}(r_{ij})& B^\prime_{m}(r_{ik}) B_{n}(r_{jk})\frac{\partial r_{ik}}{\partial R_{a,\ell}} + \nonumber \\
    B_{l}(r_{ij})& B_{m}(r_{ik}) B^\prime_{n}(r_{jk})\frac{\partial r_{jk}}{\partial R_{a,\ell}} \bigg)
    \label{eq:atomic_force}
\end{align}
with
\begin{equation}
    \frac{\partial r_{ij}}{\partial R_{a,\ell}} =\frac{( \delta _{aj} -\delta _{ai})( R_{j,\ell} -R_{i,\ell})}{r_{ij}},
\label{eq:direction_cosine}
\end{equation}
\ed{where $l$ is the Cartesian coordinate.}

\Cref{fig:workflo}B illustrates, for the two-body potential, that the choice of the cubic B-spline basis results in a smooth, continuous first derivative comprised of quadratic B-splines and a continuous second derivative of linear B-splines. 

\subsection{Regularized least-squares optimization with energies and forces}

During the fitting procedure, we optimize all spline coefficients $c$ simultaneously with the regularized linear least-squares method. Given atomic configurations $\mathcal{S}$, energies~$\mathcal{E}$, and forces~$\mathcal{F}$, we fit the potential energy function~$E$ of \cref{eq:potential} by minimizing the loss function
\ed{
\begin{align}
\label{eq:loss_function}
L & = \frac{\kappa}{\sigma_\mathcal{E}^2 |\mathcal{E}|} \sum_{s \in \mathcal{S}}(E(s)-\mathcal{E}_s)^2 \nonumber \\
& + \frac{1-\kappa}{\sigma_\mathcal{F}^2|\mathcal{F}|} \sum_{s \in \mathcal{S}}({-\nabla E(s)}-\mathcal{F}_s)^2 \nonumber \\ 
& + \lambda_1 \sum_{n}^K c_n^2 
+ \lambda_2 \sum_{n}^K (c_{n} - 2 c_{n+1} + c_{n+2})^2,
\end{align}
}
\ed{where the second sum is taken over force components. Here, $\kappa \in [0, 1]$ is a weighting parameter that controls the relative contributions between energy and force-component residuals, $|\mathcal{E}|$ and $|\mathcal{F}|$ are the number of energy and force observations in the training set,} and $\sigma_\mathcal{E}$ and $\sigma_\mathcal{F}$ are the sample standard deviations of energies and force components across the training set. This normalization yields dimensionless residuals, allowing $\kappa$ to balance the relative contributions from energies and forces, independently of the size and variance of the training energies and forces. 

The minimization of $L$ with respect to the spline coefficients $\bf c$ is a linear least-squares problem with Tikhonov regularization and solution
\ed{
\begin{equation}
    {\bf c}=({\bf X}^T{\bf X}+ \lambda_1 {\bf I} + \lambda_2 {\bf D}_2^T{\bf D}_2)^{-1} {\bf X}^T {\bf y},
\end{equation}
where ${\bf I}$ is the identity matrix, ${\bf y}$ contains energies and forces, and each element of $\bf X$ is the sum of B-spline values taken over all relevant pair distances in each configuration (rows) for each basis function (columns). Subsets of columns correspond to different body orders. Similarly, for multi-component systems, subsets of columns correspond to different chemical interactions.} When forces are included in $\bf y$, the corresponding rows of $\bf X$ are generated according to~\cref{eq:atomic_force,eq:direction_cosine}. This optimization problem is strongly convex, allowing for an efficient and deterministic solution with LU decomposition.

The used Tikhonov regularization controls the 
magnitude of spline coefficients $\bf c$ through the parameter $\lambda_1$ via the ridge penalty and the 
curvature and local smoothness across adjacent spline coefficients through $\lambda_2$ via the difference penalty \cite{Whittaker1922, Eilers1996}. For the two-body case, $D_2$ is given by
\ed{
\begin{equation}
{\bf D}_2 =
\begin{pmatrix}

1 & -2 & 1 &  &  0 \\
 & \ddots & \ddots & \ddots &  \\
 0 &  & 1 & -2 & 1

\label{eq:difference_penalty}
\end{pmatrix}.
\end{equation}
}

For higher-order potential terms, the difference penalty affects the spline coefficients that are adjacent in each dimension of a tensor product spline. This difference penalty is related to a penalty on the integral of the squared second derivative of the potential~\cite{Schoenberg1964, Reinsch1967}, used in the aPIP potentials \cite{VanderOord2019}.
However, the difference penalty is less complex because the dimensionality of the corresponding regularization problem is simply the number of basis functions $K$~\cite{Eilers1996}.

\Cref{fig:workflo}C compares optimized $\text{UF}_2$, LJ, and Morse potentials for tungsten. Although the three curves have similar minima and behavior for greater pair distances $r$, the $\text{UF}_2$ potential exhibits additional inflection points. We attribute the ability of the $\text{UF}_2$ potential to reproduce the properties of a bcc metal, a traditionally difficult task for pair potentials, to its flexible functional form. 

\ed{
The fitting of a UF potential maps energy and force data onto effective two- and three-body terms, as shown in \Cref{fig:interpret} for the tungsten dataset. Both terms can be visualized directly, providing interpretability by disentangling contributions to the interatomic interactions. The inspection of minima, repulsive and attractive contributions, and inflection points enables insight into the chemical bonding characteristics of the material. This straightforward and visual analysis makes the UF potentials more directly interpretable than most MLPs.
}


\begin{figure}[tb]
\includegraphics[width=\columnwidth]{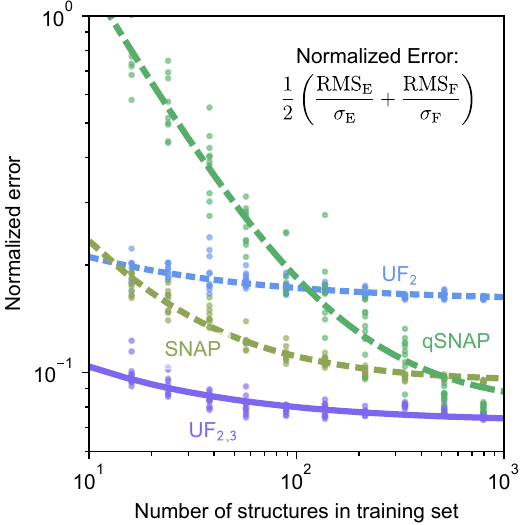}
\caption{
\emph{Learning curves for UF and SNAP potentials fit to bcc tungsten.}
Shown are out-of-sample normalized prediction errors (dots) for training sets of increasing size, with five repetitions per size.
Fitted curves (lines) are soft-plus functions that capture both the initial linear slope in log-log space and the observed saturation.
\ed{Normalized error includes energy and force error contributions weighted by the respective standard deviations of energies and forces in the training set, respectively.}
Simpler potentials saturate earlier, but with higher error than more complex potentials (UF$_2$\ vs.\ SNAP; UF$_2$\ vs.\ UF$_{2,3}$; SNAP\ vs.\ QSNAP), outperforming them when training data is limited.}
\label{fig:learning}
\end{figure}

\subsection{Convergence of error in energy and force predictions}

UF potentials should exactly reproduce any two- and three-body reference potential by construction, given sufficiently many basis functions and training data. To establish baseline functionality, we fit the $\text{UF}_2$ potential to energies and forces from the LJ potential for elemental tungsten, and the $\text{UF}_{2,3}$ potential to energies and forces from the Stillinger-Weber potential on elemental silicon, which they both reproduce with negligible error (see the Supplemental Information for learning curves and details).

To assess the accuracy of UF potentials, we measure their ability to predict DFT energies and forces in tungsten as a function of the amount of training data.
\Cref{fig:learning} compares learning curves for the two-body $\text{UF}_2$, two- and three-body $\text{UF}_{2,3}$, SNAP, and qSNAP potentials. To quantify prediction performance, we use the root-mean-squared-error (RMSE) on randomly sampled hold-out test sets. In this we ensured that each test set contained all available configuration types (see \cref{sec:data}). 
Each learning curve is fit with a softplus function $\ln(1 + e^n)$, where $n$ is the number of training data. This function captures both the initial linear slope in $\log$-$\log$ space and the observed saturation due to the models' finite complexity.

Of the four models, the $\text{UF}_2$ potential, using 28 basis functions, converges earliest. The SNAP potential, using 56  basis functions based on hyperspherical harmonics, converges slightly later with lower errors. The qSNAP potential, using 496 basis functions including quadratic terms, converges last, with modest improvements in error over SNAP. Finally, the $\text{UF}_{2,3}$ potential, using 915 basis functions, is comparable to SNAP in convergence speed with lower errors in energies and similar errors in forces. \ed{Separate energy and force learning curves are included in Fig. S5. in the supplemental information.}

The convergence speed is related to both the number of basis functions and the complexity of the many-body interactions. This indicates that the simpler $\text{UF}_2$ potential may be more suitable than other MLPs when data is scarce. 

\begin{figure*}[!tb]
\includegraphics[width=\textwidth]{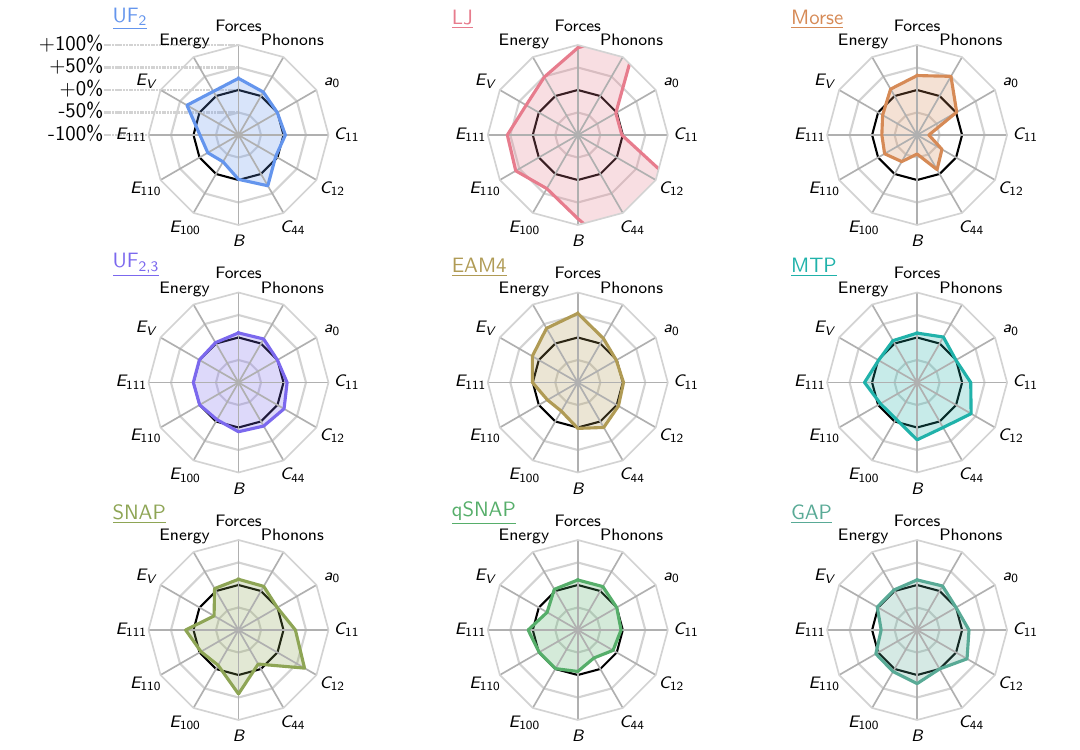}
\caption{
\emph{Performance for derived quantities} of seven potentials relative to the DFT reference \ed{for bcc tungsten}. The solid black line in each spider plot indicates zero error. Energy, force, and phonon spectra error are percent RMSE normalized by the sample standard deviation of the reference values. Other errors are percentage errors. The $\text{UF}_2$ potential achieves an accuracy approaching that of SNAP and qSNAP, while the $\text{UF}_{2,3}$ potential achieves an accuracy comparable to \ed{MTP and} GAP.
}
\label{fig:spider}
\end{figure*}


\subsection{Validation with derived quantities}
\label{results:validation}

\begin{figure}[htb]
\includegraphics[width=\columnwidth]{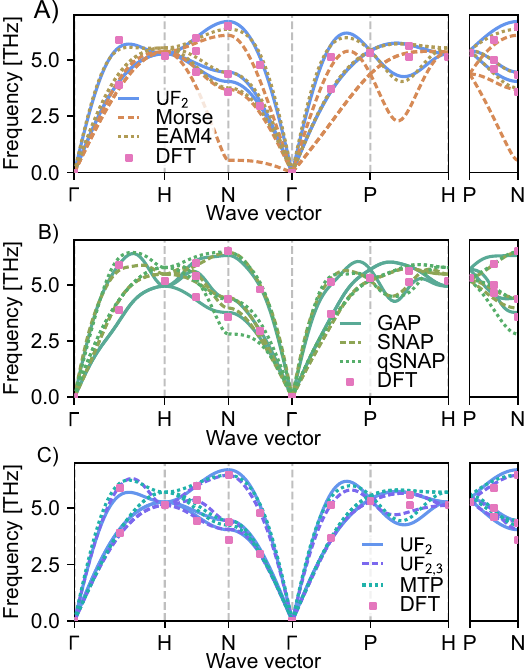}
\caption{\emph{Phonon dispersion curves \ed{for bcc tungsten.}} 
A) The $\text{UF}_2$ potential outperforms empirical potentials (Morse, EAM4) in reproducing the reference phonon frequencies (pink squares). The LJ curve, omitted, exhibits large oscillations and negative phonon frequencies. B) SNAP and qSNAP have similar phonon frequency errors to the $\text{UF}_2$ and EAM4 potentials. C) The addition of three-body interactions allows the $\text{UF}_{2,3}$ potential to approach the performance of the MTP and GAP potentials.
}
\label{fig:phonons}
\end{figure}

To benchmark performance for applications, we computed several derived quantities that were not included in the fit, such as the phonon spectrum and melting temperature, using each potential. \Cref{fig:spider} shows the relative errors in 12 quantities: energy, forces, phonon frequencies, lattice constant $a_0$, elastic constants $C_{11}$, $C_{12}$, and $C_{14}$, bulk modulus $B$, surface energies $E_{100}$, $E_{110}$, $E_{111}$, and vacancy formation energy $E_V$. Energy, force, and phonon predictions are displayed as percent RMSE, normalized by the sample standard deviation of the reference values. The remaining scalar quantities are displayed as percentage errors and tabulated in the Supplemental Information. Surface, vacancy, and strained bcc configurations were included in the training set. Hence $C_{11}$, $C_{12}$, $C_{14}$, $E_{100}$, $E_{110}$, $E_{111}$, and $E_V$ are not measures of extrapolation.

Despite its low computational cost, the UF$_2$ pair potential exhibits errors comparable to those of more complex potentials, such as SNAP and qSNAP. In contrast, the LJ and Morse potentials severely overpredict and underpredict most properties, respectively. One source of error in pair potentials, including the UF$_2$ potential, arises due to deviation from the Cauchy relations in materials. The Cauchy relations are constraints between elastic constants that hold if atoms only interact via central forces, e.g., pair potentials, and every atom is a center of inversion, such as in bcc tungsten~\cite{stakgold1950cauchy}. The Cauchy relation is $C_{12}=C_{44}$ for fcc and bcc lattices. Nobel gas crystals nearly fulfill this condition, but significant deviations occur for most other crystals. Hence, the errors in $C_{12}$ and $C_{44}$ are larger for pair potentials, where they are constrained to be equal, than for models with many-body terms. This limitation in modeling the elastic response may hinder the prediction of mechanical properties and defect-related quantities~\cite{Ziegenhain2009}.

As an example of an empirical potential used in practice, we selected the EAM4 potential~\cite{Marinica2013} for all benchmarks. The EAM4 model, one of four tungsten models developed by Marinica et al., accurately reproduces the Peierls energy barrier and dislocation core energy~\cite{Marinica2013}. The EAM4 potential was fitted to materials' properties such as those in \Cref{fig:spider} and, hence, exhibits reasonably low error.

All potentials in \cref{fig:pareto,fig:spider}, except EAM4, were fit using the same training set of 1\,939 configurations. While this training set is realistic in both size and diversity, based on their complexity the SNAP, qSNAP, \ed{MTP}, and GAP models may achieve even lower errors with a more extensive training set and larger basis set.

In this work, the cut-off radius for inter-atomic interactions was set to 5.5\;\AA{} for all potentials except for EAM4. Additional hyperparameters for the $\text{UF}_{2}$, $\text{UF}_{2,3}$, SNAP, qSNAP, and GAP basis sets are tabulated in the Supplemental Information. For $\text{UF}_{2,3}$, a separate smaller cut-off radius of 4.25\;\AA{} was used for three-body interactions. This choice was motivated in part by precedence in other two- and three-body potentials such as the modified embedded-atom potentials \cite{Hennig2008} and in part by speed: The smaller cut-off radius results in ten times fewer three-body interactions and corresponding speed-up.

For the bcc tungsten system, the $\text{UF}_{2,3}$ potential approaches the accuracy of the \ed{MTP and GAP potentials}. Adding the three-body interactions eliminates the error associated with the Cauchy discrepancy, improving the elastic constant predictions. The $\text{UF}_{2,3}$ potential also achieves lower errors for the surface and vacancy formation energies than the $\text{UF}_{2}$ pair potential, underscoring the value of including three-body interactions.

\begin{table}[tb]
\ed{
\begin{tabular}{@{}lcccccc@{}}
    \toprule
 &
  \multicolumn{1}{p{1.2cm}}{\centering Relative\\Comp.\\Cost} &
  \multicolumn{1}{p{1.2cm}}{\centering Melting\\Temp.\\(K)} &
  \multicolumn{1}{p{1.5cm}}{\centering Energy\\RMSE\\(eV/atom)} &
  \multicolumn{1}{p{1.3cm}}{\centering Force\\RMSE\\(eV/\AA)} &
  \multicolumn{1}{p{1.2cm}}{\centering Phonon\\RMSE (THz)}
  \\ \midrule
    DFT~\cite{Wang2011} &  & $3465\pm 105$ &       &       &      \\
    LJ               & 1        & $5695 \pm 90$ & 0.110 & 1.400 & 3.914\\
    Morse            & 1.55     & $2681 \pm 45$ & 0.040 & 0.480 & 1.140\\
    $\text{UF}_2$    & 0.79     & $3850 \pm 68$ & 0.027 & 0.387 & 0.230\\
    EAM4             & 2.92     & $4573\pm 78$ & 0.088 & 0.803 & 0.301\\
    $\text{UF}_{2,3}$& 10.5\footnotemark & $3651\pm 31$    & 0.005 & 0.152 & 0.263\\
    MTP              &  45.7     & $3961\pm 82$ & 0.017 & 0.146 & 0.376 \\
    qSNAP            & 145      & -            & 0.010 & 0.167 & 0.256\\
    SNAP             & 443      & $3136\pm 63$ & 0.014 & 0.189 & 0.270\\
    GAP              & 3070     & $3141\pm 54$ & 0.006 & 0.169 & 0.291\\
    \bottomrule
\end{tabular}
}
\footnotemark[1] See \cref{sec:methods:calcs}.

\caption{
\emph{Melting temperature predictions alongside energy, force, and phonon frequency benchmarks \ed{for bcc tungsten.}} See \cref{results:validation} and \cref{sec:methods:calcs} for details. 
Compared to LJ, Morse, and EAM4, the $\text{UF}_2$ potential achieves a low error in the melting temperature for a similar cost. \ed{The  $\text{UF}_{2,3}$ potential prediction is even closer to the DFT reference at the cost of one order of magnitude in speed. MTP,} SNAP, qSNAP, and GAP require one to three orders of magnitude more computational resources for the same large-scale simulation.
}
\label{table:comparison}
\end{table}

\Cref{fig:phonons} compares the calculated phonon spectra of the various MLPs with the DFT reference values~\cite{Szlachta2014}. Perhaps surprisingly, the $\text{UF}_2$ pair potential is comparable in error to the potentials with many-body terms. In contrast, the Morse pair potential is rather inaccurate and the LJ pair potential, not shown, yields a phonon spectrum with imaginary frequency and large frequency oscillations. \Cref{table:comparison} summarizes the RMSE of the phonon frequencies computed across the 26 DFT reference values. As in other calculated properties, the addition of three-body interactions in the $\text{UF}_{2,3}$ potential significantly improves the agreement with the DFT reference compared to the $\text{UF}_2$ pair potential and even surpasses the other computationally more expensive MLPs.

As an example for a practical, large-scale calculation, we predict the melting temperature of tungsten (see \cref{sec:methods:calcs} for computational details). Since the training set of the MLPs does not include liquid configurations, the melting point predictions measure the models' extrapolative capacity. \Cref{table:comparison} compares the predicted melting temperatures of the MLPs \ed{to the \emph{ab-initio} reference value of 3465\;K~\cite{Wang2011}.} 
We observe that the other pair potentials are limited in their predictive capabilities while the $\text{UF}_2$ potential is comparable in accuracy to the more complex potentials, which yield reasonable predictions. The qSNAP melting calculation failed due to numerical instability at higher temperatures, which is known to occur in high-dimensional potentials. \cite{lqb2021q} Bonds break at higher temperatures, leading to many local atomic configurations that are underrepresented in the training set. We expect that training the qSNAP potential on a suitable, more extensive dataset would remove this instability. The extrapolative capacity of the $\text{UF}_2$ and $\text{UF}_{2,3}$ potentials in melting-temperature simulations illustrates that the UF potentials can provide an accurate description with only a moderately sized training dataset, indicating their usefulness for materials simulations with limited reference data.

\section{Summary}

We developed and implemented a machine-learning potential that is fast to train and evaluate, provides an interpretable form, is extendable to higher-order interactions, and accurately describes materials even for comparably sparse training sets. The approach is based on an effective many-body expansion and utilizes a flexible B-spline basis. The resulting regularized linear least-squares optimization problem is strongly convex, significantly reducing the computational requirements.

For the example of elemental tungsten, the $\text{UF}_2$ pair potential produces energy, force, and property predictions rivaling those of SNAP and qSNAP while matching the cost of the Morse potential, corresponding to a reduction in computational cost by two orders of magnitude. Despite the intrinsic limitations of the pair potential in capturing physics, we find that the $\text{UF}_2$ pair potential yields reasonable predictions in property benchmarks, such as for elastic constants, phonons, surface energies, and melting temperature.
The $\text{UF}_{2,3}$ potential, which accounts for three-body interactions, approaches the accuracy of \ed{MTP and} GAP while reducing computational cost by one to three orders of magnitude.


\ed{
The rapidly increasing number of MLPs, from ultra-fast linear models to graph neural network potentials, highlights the trade-offs between computational efficiency, robustness, and model capacity. Complex, high-dimensional MLPs are expected to yield higher accuracy for complicated systems at the price of greatly increased computational cost and data requirements. The UF approach yields potentials that are fast and robust, at the price of reduced flexibility and possibly greater errors for complex systems. Future work on UF potentials will explore the use of active learning for increased robustness and data efficiency as well as the addition of the four-body term, which is necessary for modeling dihedral angles that are critical to describing organic molecules and protein structures.
}
The software for fitting UF potentials and exporting LAMMPS-compatible tables is freely available in our Github repository~\cite{xie2021uf}.

\section{Methods}
\label{sec:methods}

\subsection{Data}\label{sec:data}

To compare the UF potential against existing potentials we use a dataset by Szlachta et al. \cite{Szlachta2014} which has been used before to benchmark the GAP \cite{Szlachta2014}, SNAP \cite{Wood2017}, and aPIP \cite{VanderOord2019} potentials.
This dataset of 9\,693 tungsten configurations includes body-centered cubic (bcc) primitive cells, bulk snapshots from molecular dynamics, surfaces, vacancies, gamma surfaces,  gamma surface vacancies, and dislocation quadrupoles.
Energies, forces, and stresses in the dataset were computed using density functional theory (DFT) with the Perdew-Burke-Ernzerhof (PBE) \cite{Perdew1996} functional.

\subsection{B-spline basis}

The UF potential uses natural cubic B-splines: Their first derivative is continuous and smooth, while their second derivative is continuous. They are natural, rather than clamped, in that their second derivative is zero at the boundary conditions. These properties, which are critical for accurately reproducing forces and stresses,  motivated our choice of basis set. Other B-spline schemes have been explored for interpolation in empirical potential development. Wen et al.{}~\cite{Wen2015} discuss clamped and Hermite splines as well as quartic and quintic splines. The natural cubic spline is sufficient except when computing properties that rely on the third and fourth derivative of the potential, such as thermal expansion and finite-temperature elastic constants.

Uniform spacing of knots is a reasonable choice in many cases. However, the user may adjust the density of knots to control resolution in regions of interest. Due to compact support in this basis, each pair-distance energy requires the evaluation of exactly four B-splines. Hence, the potential's speed scales with neither the number of knots nor the number of basis functions. 

On the other hand, the minimum distance between knots limits the maximum curvature of the function. The optimum density of knots thus depends on the quality of the training set available. Underfitting or overfitting may arise from insufficient or excessive knot density, respectively. Based on convergence tests (supplemental information), we fit UF potentials in this work using 25 uniformly spaced knots.

By construction, each spline coefficient influences the overall function across five adjacent knots. As a result, the user can tune the shape of the potential further according to additional constraints. For instance, soft-core and smooth-cutoff requirements may be satisfied by adjusting spline coefficients at the ends. In this work, we ensure that the potential and its first derivative follow a smooth cutoff at $r_{ij} = t_K$ by setting the last three coefficients to 0.

\subsection{Models}

In this work, we partitioned the data using a random split of 20\% training and 80\% testing data. The testing set was used to evaluate the root-mean-square error (RMSE) in energies and forces, as reported in \cref{table:comparison} and \cref{fig:pareto}. The LJ and Morse potentials were optimized using the BFGS algorithm as implemented in the SciPy library~\cite{Virtanen2020}. The SNAP and qSNAP potentials were retrained using the MAterials Machine Learning (MAML) package~\cite{PingOng2019}. The GAP potential was retrained using the QUantum mechanics and Interatomic Potentials (QUIP) package~\cite{Bartok2013, Bartok2015}. We used the EAM4 potential, obtained from the NIST Interatomic Potential Repository \cite{Becker2013, Hale2018}, without modification. \ed{The MTP potential was fit using the MLIP package~\cite{Novikov2021}.}

\ed{The size of the training set was selected to represent common, data-scarce scenarios.}
With a larger training set, the \ed{MTP,} SNAP, qSNAP, and GAP models would likely produce better predictions. We refer the reader to the original works and existing benchmarks \cite{Zuo2020,Vita2021} for details regarding convergence with training examples and model complexity.
\subsection{Calculations}\label{sec:methods:calcs}

All potentials in \cref{fig:pareto} were benchmarked using one thread on an AMD EPYC 7702 Rome (2.0 GHz) CPU.
Computational cost measurements for each potential are reported as the average over ten simulations. 

We use two methods to estimate the computational cost of the $\text{UF}_{2,3}$ potential and present both values in \cref{fig:pareto} and \cref{table:comparison}. The lower estimate, 0.76 ms/step, is computed by multiplying the computational cost of $\text{UF}_2$ by the ratio of floating point operations used by $\text{UF}_{2,3}$ and $\text{UF}_2$. The higher estimate, 2.03 ms/step, is computed using the ratio of speeds, in the Python implementation, multiplied by the reported cost of $\text{UF}_2$. We show performance bounds instead of the current $\text{UF}_{2,3}$ implementation's computational cost because it has not been fully optimized yet.

Elastic constants were evaluated using the Elastic python package \cite{Jochym2018}. Phonon spectra were evaluated using the Phonopy python package \cite{Togo2015}. Melting temperatures were calculated in LAMMPS using the two-phase method and a timestep of 1 fs. The initial system, a $16 \times 8 \times 8$ bcc supercell (2\,048 atoms), was equilibrated at a selected temperature for 40\,000 timesteps. Next, the solid-phase atoms were fixed while the liquid-phase atoms were heated to 5\,000\;K and cooled back to the initial temperature over 80\,000 timesteps. Finally, the system was equilibrated over 200\,000 steps using the isoenthalpic-isobaric (NPH) ensemble. If the final configuration did not contain both phases, the procedure was repeated with a different initial temperature. Reported melting temperatures were computed by taking the average over the final 100\,000 steps.

\section*{Code availability}

Code for fitting UF$_2$ and UF$_3$ potentials, as well as exporting LAMMPS-compatible tables is freely available in our open-source ``Ultra-Fast Force Fields'' GitHub repository~\cite{xie2021uf}. The $\text{UF}_2$ potential is natively supported by LAMMPS, GROMACS, and other molecular dynamics suites that can construct potentials from interpolation tables. In LAMMPS, the pair style ``table" is available for execution on both CPU and GPUs, enabling the $\text{UF}_2$ potential to benefit from various computer architectures. \ed{A LAMMPS package for using $\text{UF}_{2,3}$ potentials is also available in the repository.}

\vspace{1em}

\ed{
\section*{Data availability}

Example notebooks, LAMMPS input files, and parameters for all potentials fit in this work are available in the GitHub repository~\cite{xie2021uf}. The tungsten and silicon datasets are publicly available~\cite{data_w,data_si}.
}

\section*{Acknowledgments}

SX and RGH were supported by the United States Department of Energy, under contract number DE-SC0020385. RGH was supported by the U.\ S.\ National Science Foundation under contract number DMR 2118718. MR acknowledges partial support by the European Centre of Excellence in Exascale Computing TREX---Targeting Real Chemical Accuracy at the Exascale; this project has received funding from the European Union's Horizon 2020 Research and Innovation program under Grant Agreement No.~952165. Part of the research was performed while the authors visited the Institute for Pure and Applied Mathematics (IPAM), which is supported by the National Science Foundation (Grant No. DMS 1440415). Computational resources were provided by the University of Florida Research Computing Center.

We thank Ajinkya Hire for the implementation of UF potentials in LAMMPS and Alexander Shapeev for fitting the MTP potentials. 
We thank 
Thomas Bischoff, 
Jason Gibson, 
Bastian J{\"a}ckl, 
Hendrik Kra{\ss}, 
Ming Li,
Johannes Margraf,
Paul-Rene Mayer,
Pawan Prakash, 
Robert Schmid,
and Benjamin Walls 
for testing of and contributing to the UF implementation.


\section*{Author Contributions}

All authors contributed extensively to the work presented in this paper. SRX, MR, and RGH jointly develped the methodology. SRX implemented the algorithm and performed the model training and analysis. SRX, MR, and RGH wrote the manuscript.

\section*{Competing Interests statement}
The Authors declare no Competing Financial or Non- Financial Interests.

\bibliography{uf_references}

\begin{thebibliography}{10}
\expandafter\ifx\csname url\endcsname\relax
  \def\url#1{\texttt{#1}}\fi
\expandafter\ifx\csname urlprefix\endcsname\relax\def\urlprefix{URL }\fi
\providecommand{\bibinfo}[2]{#2}
\providecommand{\eprint}[2][]{\url{#2}}

\bibitem{Deringer2019}
\bibinfo{author}{Deringer, V.~L.}, \bibinfo{author}{Caro, M.~A.} \&
  \bibinfo{author}{Cs{\'a}nyi, G.}
\newblock \bibinfo{title}{Machine learning interatomic potentials as emerging
  tools for materials science}.
\newblock \emph{\bibinfo{journal}{Adv. Mater.}} \textbf{\bibinfo{volume}{31}},
  \bibinfo{pages}{1902765} (\bibinfo{year}{2019}).

\bibitem{Langer2022}
\bibinfo{author}{Langer, M.~F.}, \bibinfo{author}{Goe{\ss}mann, A.} \&
  \bibinfo{author}{Rupp, M.}
\newblock \bibinfo{title}{Representations of molecules and materials for
  interpolation of quantum-mechanical simulations via machine learning}.
\newblock \emph{\bibinfo{journal}{npj Comput. Mater.}}
  \textbf{\bibinfo{volume}{8}}, \bibinfo{pages}{41} (\bibinfo{year}{2022}).

\bibitem{Miksch2021}
\bibinfo{author}{Miksch, A.~M.}, \bibinfo{author}{Morawietz, T.},
  \bibinfo{author}{K{\"a}stner, J.}, \bibinfo{author}{Urban, A.} \&
  \bibinfo{author}{Artrith, N.}
\newblock \bibinfo{title}{Strategies for the construction of machine-learning
  potentials for accurate and efficient atomic-scale simulations}.
\newblock \emph{\bibinfo{journal}{Mach. Learn. Sci. Tech.}}
  \textbf{\bibinfo{volume}{2}}, \bibinfo{pages}{031001} (\bibinfo{year}{2021}).

\bibitem{Friederich2021}
\bibinfo{author}{Friederich, P.}, \bibinfo{author}{H\"{a}se, F.},
  \bibinfo{author}{Proppe, J.} \& \bibinfo{author}{Aspuru-Guzik, A.}
\newblock \bibinfo{title}{Machine-learned potentials for next-generation matter
  simulations}.
\newblock \emph{\bibinfo{journal}{Nat. Mater.}} \textbf{\bibinfo{volume}{20}},
  \bibinfo{pages}{750--761} (\bibinfo{year}{2021}).

\bibitem{Drautz2004}
\bibinfo{author}{Drautz, R.}, \bibinfo{author}{Fähnle, M.} \&
  \bibinfo{author}{Sanchez, J.~M.}
\newblock \bibinfo{title}{General relations between many-body potentials and
  cluster expansions in multicomponent systems}.
\newblock \emph{\bibinfo{journal}{J. Phys.: Condens. Matter}}
  \textbf{\bibinfo{volume}{16}}, \bibinfo{pages}{3843--3852}
  (\bibinfo{year}{2004}).

\bibitem{rapaport2004}
\bibinfo{author}{Rapaport, D.}
\newblock \emph{\bibinfo{title}{The Art of Molecular Dynamics Simulation}}
  (\bibinfo{publisher}{Cambridge University Press}, \bibinfo{year}{2004}).

\bibitem{Martinez2013}
\bibinfo{author}{Martinez, J.~A.}, \bibinfo{author}{Yilmaz, D.~E.},
  \bibinfo{author}{Liang, T.}, \bibinfo{author}{Sinnott, S.~B.} \&
  \bibinfo{author}{Phillpot, S.~R.}
\newblock \bibinfo{title}{Fitting empirical potentials: Challenges and
  methodologies}.
\newblock \emph{\bibinfo{journal}{Curr. Opin. Solid State Mater. Sci}}
  \textbf{\bibinfo{volume}{17}}, \bibinfo{pages}{263--270}
  (\bibinfo{year}{2013}).

\bibitem{Ragasa2019}
\bibinfo{author}{Ragasa, E.~J.}, \bibinfo{author}{O’Brien, C.~J.},
  \bibinfo{author}{Hennig, R.~G.}, \bibinfo{author}{Foiles, S.~M.} \&
  \bibinfo{author}{Phillpot, S.~R.}
\newblock \bibinfo{title}{Multi-objective optimization of interatomic
  potentials with application to {MgO}}.
\newblock \emph{\bibinfo{journal}{Model. Simul. Mater. Sci. Eng.}}
  \textbf{\bibinfo{volume}{27}}, \bibinfo{pages}{074007}
  (\bibinfo{year}{2019}).

\bibitem{Jones1924}
\bibinfo{author}{Jones, J.~E.}
\newblock \bibinfo{title}{On the determination of molecular {fields.---I.} from
  the variation of the viscosity of a gas with temperature}.
\newblock \emph{\bibinfo{journal}{Proc. R. Soc. Lond. A}}
  \textbf{\bibinfo{volume}{106}}, \bibinfo{pages}{441--462}
  (\bibinfo{year}{1924}).

\bibitem{Morse1929}
\bibinfo{author}{Morse, P.~M.}
\newblock \bibinfo{title}{Diatomic molecules according to the wave mechanics.
  {II.} vibrational levels}.
\newblock \emph{\bibinfo{journal}{Phys. Rev.}} \textbf{\bibinfo{volume}{34}},
  \bibinfo{pages}{57--64} (\bibinfo{year}{1929}).

\bibitem{Daw1984}
\bibinfo{author}{Daw, M.~S.} \& \bibinfo{author}{Baskes, M.~I.}
\newblock \bibinfo{title}{Embedded-atom method: Derivation and application to
  impurities, surfaces, and other defects in metals}.
\newblock \emph{\bibinfo{journal}{Phys. Rev. B}} \textbf{\bibinfo{volume}{29}},
  \bibinfo{pages}{6443--6453} (\bibinfo{year}{1984}).

\bibitem{mgcc2021q}
\bibinfo{author}{Musil, F.} \emph{et~al.}
\newblock \bibinfo{title}{Physics-inspired structural representations for
  molecules and materials}.
\newblock \emph{\bibinfo{journal}{Chem. Rev.}} \textbf{\bibinfo{volume}{121}},
  \bibinfo{pages}{9759--9815} (\bibinfo{year}{2021}).

\bibitem{dbcc2021q}
\bibinfo{author}{Deringer, V.~L.} \emph{et~al.}
\newblock \bibinfo{title}{{G}aussian process regression for materials and
  molecules}.
\newblock \emph{\bibinfo{journal}{Chem. Rev.}} \textbf{\bibinfo{volume}{121}},
  \bibinfo{pages}{10073--10141} (\bibinfo{year}{2021}).

\bibitem{hl2021q}
\bibinfo{author}{Huang, B.} \& \bibinfo{author}{von Lilienfeld, O.~A.}
\newblock \bibinfo{title}{Ab initio machine learning in chemical compound
  space}.
\newblock \emph{\bibinfo{journal}{Chem. Rev.}} \textbf{\bibinfo{volume}{121}},
  \bibinfo{pages}{10001--10036} (\bibinfo{year}{2021}).

\bibitem{b2021q}
\bibinfo{author}{Behler, J.}
\newblock \bibinfo{title}{Four generations of high-dimensional neural network
  potentials}.
\newblock \emph{\bibinfo{journal}{Chem. Rev.}} \textbf{\bibinfo{volume}{121}},
  \bibinfo{pages}{10037--10072} (\bibinfo{year}{2021}).

\bibitem{uctm2021q}
\bibinfo{author}{Unke, O.~T.} \emph{et~al.}
\newblock \bibinfo{title}{Machine learning force fields}.
\newblock \emph{\bibinfo{journal}{Chem. Rev.}} \textbf{\bibinfo{volume}{121}},
  \bibinfo{pages}{10142--10186} (\bibinfo{year}{2021}).

\bibitem{Parsaeifard2020}
\bibinfo{author}{Parsaeifard, B.} \emph{et~al.}
\newblock \bibinfo{title}{An assessment of the structural resolution of various
  fingerprints commonly used in machine learning}.
\newblock \emph{\bibinfo{journal}{Mach. Learn.: Sci. Technol.}}
  \textbf{\bibinfo{volume}{2}}, \bibinfo{pages}{015018} (\bibinfo{year}{2021}).

\bibitem{Zuo2020}
\bibinfo{author}{Zuo, Y.} \emph{et~al.}
\newblock \bibinfo{title}{Performance and cost assessment of machine learning
  interatomic potentials}.
\newblock \emph{\bibinfo{journal}{J. Phys. Chem. A}}
  \textbf{\bibinfo{volume}{124}}, \bibinfo{pages}{731--745}
  (\bibinfo{year}{2020}).

\bibitem{Lysogorskiy2021}
\bibinfo{author}{Lysogorskiy, Y.} \emph{et~al.}
\newblock \bibinfo{title}{Performant implementation of the atomic cluster
  expansion ({PACE}) and application to copper and silicon}.
\newblock \emph{\bibinfo{journal}{npj Comput. Mater.}}
  \textbf{\bibinfo{volume}{7}} (\bibinfo{year}{2021}).

\bibitem{Kovacs2021}
\bibinfo{author}{Kov{\'a}cs, D.~P.} \emph{et~al.}
\newblock \bibinfo{title}{Linear atomic cluster expansion force fields for
  organic molecules: Beyond {RMSE}}.
\newblock \emph{\bibinfo{journal}{J. Chem. Theor. Comput.}}
  \textbf{\bibinfo{volume}{17}}, \bibinfo{pages}{7696--7711}
  (\bibinfo{year}{2021}).

\bibitem{Thompson2015}
\bibinfo{author}{Thompson, A.}, \bibinfo{author}{Swiler, L.},
  \bibinfo{author}{Trott, C.}, \bibinfo{author}{Foiles, S.} \&
  \bibinfo{author}{Tucker, G.}
\newblock \bibinfo{title}{Spectral neighbor analysis method for automated
  generation of quantum-accurate interatomic potentials}.
\newblock \emph{\bibinfo{journal}{J. Comput. Phys.}}
  \textbf{\bibinfo{volume}{285}}, \bibinfo{pages}{316--330}
  (\bibinfo{year}{2015}).

\bibitem{Wood2018}
\bibinfo{author}{Wood, M.~A.} \& \bibinfo{author}{Thompson, A.~P.}
\newblock \bibinfo{title}{Extending the accuracy of the {SNAP} interatomic
  potential form}.
\newblock \emph{\bibinfo{journal}{J. Chem. Phys.}}
  \textbf{\bibinfo{volume}{148}}, \bibinfo{pages}{241721}
  (\bibinfo{year}{2018}).

\bibitem{Bartok2009}
\bibinfo{author}{Bartók, A.~P.}, \bibinfo{author}{Payne, M.~C.},
  \bibinfo{author}{Kondor, R.} \& \bibinfo{author}{Csányi, G.}
\newblock \bibinfo{title}{{Gaussian} approximation potentials: {The} accuracy
  of quantum mechanics, without the electrons}.
\newblock \emph{\bibinfo{journal}{Phys. Rev. Lett.}}
  \textbf{\bibinfo{volume}{104}} (\bibinfo{year}{2010}).

\bibitem{Shapeev2016}
\bibinfo{author}{Shapeev, A.~V.}
\newblock \bibinfo{title}{Moment tensor potentials: {A} class of systematically
  improvable interatomic potentials}.
\newblock \emph{\bibinfo{journal}{Multiscale Model. Simul.}}
  \textbf{\bibinfo{volume}{14}}, \bibinfo{pages}{1153--1173}
  (\bibinfo{year}{2016}).

\bibitem{Drautz2019}
\bibinfo{author}{Drautz, R.}
\newblock \bibinfo{title}{Atomic cluster expansion for accurate and
  transferable interatomic potentials}.
\newblock \emph{\bibinfo{journal}{Phys. Rev. B}} \textbf{\bibinfo{volume}{99}},
  \bibinfo{pages}{14104} (\bibinfo{year}{2019}).

\bibitem{VanderOord2019}
\bibinfo{author}{van~der Oord, C.}, \bibinfo{author}{Dusson, G.},
  \bibinfo{author}{Cs{\'{a}}nyi, G.} \& \bibinfo{author}{Ortner, C.}
\newblock \bibinfo{title}{Regularised atomic body-ordered permutation-invariant
  polynomials for the construction of interatomic potentials}.
\newblock \emph{\bibinfo{journal}{Mach. Learn.: Sci. Technol.}}
  \textbf{\bibinfo{volume}{1}}, \bibinfo{pages}{015004} (\bibinfo{year}{2020}).

\bibitem{lindsey2017}
\bibinfo{author}{Lindsey, R.~K.}, \bibinfo{author}{Fried, L.~E.} \&
  \bibinfo{author}{Goldman, N.}
\newblock \bibinfo{title}{{ChIMES}: A force matched potential with explicit
  three-body interactions for molten carbon} \textbf{\bibinfo{volume}{13}},
  \bibinfo{pages}{6222--6229} (\bibinfo{year}{2017}).

\bibitem{Vandermause2020}
\bibinfo{author}{Vandermause, J.} \emph{et~al.}
\newblock \bibinfo{title}{On-the-fly active learning of interpretable
  {Bayesian} force fields for atomistic rare events}.
\newblock \emph{\bibinfo{journal}{npj Comput. Mater.}}
  \textbf{\bibinfo{volume}{6}} (\bibinfo{year}{2020}).

\bibitem{pozdnyakov2020}
\bibinfo{author}{Pozdnyakov, S.}, \bibinfo{author}{Oganov, A.~R.},
  \bibinfo{author}{Mazitov, A.}, \bibinfo{author}{Kruglov, I.} \&
  \bibinfo{author}{Mazhnik, E.}
\newblock \bibinfo{title}{Fast general two- and three-body interatomic
  potential}.
\newblock \emph{\bibinfo{journal}{arXiv}} \bibinfo{pages}{1910.07513}
  (\bibinfo{year}{2020}).

\bibitem{Behler2007}
\bibinfo{author}{Behler, J.} \& \bibinfo{author}{Parrinello, M.}
\newblock \bibinfo{title}{Generalized neural-network representation of
  high-dimensional potential-energy surfaces}.
\newblock \emph{\bibinfo{journal}{Phys. Rev. Lett.}}
  \textbf{\bibinfo{volume}{98}} (\bibinfo{year}{2007}).

\bibitem{Bircher2021}
\bibinfo{author}{Bircher, M.~P.}, \bibinfo{author}{Singraber, A.} \&
  \bibinfo{author}{Dellago, C.}
\newblock \bibinfo{title}{Improved description of atomic environments using
  low-cost polynomial functions with compact support}.
\newblock \emph{\bibinfo{journal}{Mach. Learn.: Sci. Technol.}}
  \textbf{\bibinfo{volume}{2}}, \bibinfo{pages}{035026} (\bibinfo{year}{2021}).

\bibitem{Vita2021}
\bibinfo{author}{Vita, J.~A.} \& \bibinfo{author}{Trinkle, D.~R.}
\newblock \bibinfo{title}{Exploring the necessary complexity of interatomic
  potentials}.
\newblock \emph{\bibinfo{journal}{Comput. Mater. Sci.}}
  \textbf{\bibinfo{volume}{200}}, \bibinfo{pages}{110752}
  (\bibinfo{year}{2021}).

\bibitem{Szlachta2014}
\bibinfo{author}{Szlachta, W.~J.}, \bibinfo{author}{Bartók, A.~P.} \&
  \bibinfo{author}{Csányi, G.}
\newblock \bibinfo{title}{Accuracy and transferability of {Gaussian}
  approximation potential models for tungsten}.
\newblock \emph{\bibinfo{journal}{Phys. Rev. B}} \textbf{\bibinfo{volume}{90}},
  \bibinfo{pages}{104108} (\bibinfo{year}{2014}).

\bibitem{xie2021uf}
\bibinfo{author}{Xie, S.} \& \bibinfo{author}{Rupp, M.}
\newblock \bibinfo{title}{Ultra fast force fields package}.
\newblock \bibinfo{howpublished}{\url {https://github.com/uf3}}
  (\bibinfo{year}{2021}).

\bibitem{kresse1996}
\bibinfo{author}{Kresse, G.} \& \bibinfo{author}{Furthm{\"u}ller, J.}
\newblock \bibinfo{title}{Efficient iterative schemes for ab initio
  total-energy calculations using a plane-wave basis set}.
\newblock \emph{\bibinfo{journal}{Phys. Rev. B}} \textbf{\bibinfo{volume}{54}},
  \bibinfo{pages}{11169} (\bibinfo{year}{1996}).

\bibitem{Plimpton1995}
\bibinfo{author}{Plimpton, S.}
\newblock \bibinfo{title}{Fast parallel algorithms for short-range molecular
  dynamics}.
\newblock \emph{\bibinfo{journal}{J. Comput. Phys.}}
  \textbf{\bibinfo{volume}{117}}, \bibinfo{pages}{1--19}
  (\bibinfo{year}{1995}).

\bibitem{tabbbcvkmnsetal2022}
\bibinfo{author}{Thompson, A.~P.} \emph{et~al.}
\newblock \bibinfo{title}{{LAMMPS}---a flexible simulation tool for
  particle-based materials modeling at the atomic, meso, and continuum scales}.
\newblock \emph{\bibinfo{journal}{Comput. Phys. Comm.}}
  \textbf{\bibinfo{volume}{271}}, \bibinfo{pages}{108171}
  (\bibinfo{year}{2022}).

\bibitem{boor2001}
\bibinfo{author}{de~Boor, C.}
\newblock \emph{\bibinfo{title}{A Practical Guide to Splines}}
  (\bibinfo{publisher}{Springer}, \bibinfo{address}{New York},
  \bibinfo{year}{1978}).

\bibitem{Runge1901}
\bibinfo{author}{Runge, C.}
\newblock \bibinfo{title}{{\"U}ber empirische {F}unktionen und die
  {I}nterpolation zwischen {\"a}quidistanten {O}rdinaten}.
\newblock \emph{\bibinfo{journal}{Z. Math. Phys.}}
  \textbf{\bibinfo{volume}{46}}, \bibinfo{pages}{224--243}
  (\bibinfo{year}{1901}).

\bibitem{Wolff1999}
\bibinfo{author}{Wolff, D.} \& \bibinfo{author}{Rudd, W.}
\newblock \bibinfo{title}{Tabulated potentials in molecular dynamics
  simulations}.
\newblock \emph{\bibinfo{journal}{Comput. Phys. Commun.}}
  \textbf{\bibinfo{volume}{120}}, \bibinfo{pages}{20--32}
  (\bibinfo{year}{1999}).

\bibitem{Wen2015}
\bibinfo{author}{Wen, M.}, \bibinfo{author}{Whalen, S.~M.},
  \bibinfo{author}{Elliott, R.~S.} \& \bibinfo{author}{Tadmor, E.~B.}
\newblock \bibinfo{title}{Interpolation effects in tabulated interatomic
  potentials}.
\newblock \emph{\bibinfo{journal}{Model. Simul. Mater. Sci. Eng.}}
  \textbf{\bibinfo{volume}{23}}, \bibinfo{pages}{074008}
  (\bibinfo{year}{2015}).

\bibitem{Hennig2008}
\bibinfo{author}{Hennig, R.}, \bibinfo{author}{Lenosky, T.},
  \bibinfo{author}{Trinkle, D.}, \bibinfo{author}{Rudin, S.} \&
  \bibinfo{author}{Wilkins, J.}
\newblock \bibinfo{title}{Classical potential describes martensitic phase
  transformations between the $\alpha$, $\beta$, and $\omega$ titanium phases}.
\newblock \emph{\bibinfo{journal}{Phys. Rev. B}} \textbf{\bibinfo{volume}{78}}
  (\bibinfo{year}{2008}).

\bibitem{Whittaker1922}
\bibinfo{author}{Whittaker, E.~T.}
\newblock \bibinfo{title}{On a new method of graduation}.
\newblock \emph{\bibinfo{journal}{Proceedings of the Edinburgh Mathematical
  Society}} \textbf{\bibinfo{volume}{41}}, \bibinfo{pages}{63–75}
  (\bibinfo{year}{1922}).

\bibitem{Eilers1996}
\bibinfo{author}{Eilers, P.~H.} \& \bibinfo{author}{Marx, B.~D.}
\newblock \bibinfo{title}{Flexible smoothing with {B}-splines and penalties}.
\newblock \emph{\bibinfo{journal}{Statist. Sci.}} \textbf{\bibinfo{volume}{11}}
  (\bibinfo{year}{1996}).

\bibitem{Schoenberg1964}
\bibinfo{author}{Schoenberg, I.~J.}
\newblock \bibinfo{title}{Spline functions and the problem of graduation}.
\newblock \emph{\bibinfo{journal}{Proc. Natl. Acad. Sci. USA}}
  \textbf{\bibinfo{volume}{52}}, \bibinfo{pages}{947--950}
  (\bibinfo{year}{1964}).

\bibitem{Reinsch1967}
\bibinfo{author}{Reinsch, C.~H.}
\newblock \bibinfo{title}{Smoothing by spline functions}.
\newblock \emph{\bibinfo{journal}{Numer. Math.}} \textbf{\bibinfo{volume}{10}},
  \bibinfo{pages}{177--183} (\bibinfo{year}{1967}).

\bibitem{stakgold1950cauchy}
\bibinfo{author}{Stakgold, I.}
\newblock \bibinfo{title}{The {C}auchy relations in a molecular theory of
  elasticity}.
\newblock \emph{\bibinfo{journal}{Q. Appl. Math.}}
  \textbf{\bibinfo{volume}{8}}, \bibinfo{pages}{169--186}
  (\bibinfo{year}{1950}).

\bibitem{Ziegenhain2009}
\bibinfo{author}{Ziegenhain, G.}, \bibinfo{author}{Hartmaier, A.} \&
  \bibinfo{author}{Urbassek, H.~M.}
\newblock \bibinfo{title}{{Pair vs many-body potentials: {Influence} on elastic
  and plastic behavior in nanoindentation of fcc metals}}.
\newblock \emph{\bibinfo{journal}{J. Mech. Phys. Solids}}
  \textbf{\bibinfo{volume}{57}}, \bibinfo{pages}{1514--1526}
  (\bibinfo{year}{2009}).

\bibitem{Marinica2013}
\bibinfo{author}{Marinica, M.-C.} \emph{et~al.}
\newblock \bibinfo{title}{Interatomic potentials for modelling radiation
  defects and dislocations in tungsten}.
\newblock \emph{\bibinfo{journal}{J. Phys.: Condens. Matter}}
  \textbf{\bibinfo{volume}{25}}, \bibinfo{pages}{395502}
  (\bibinfo{year}{2013}).

\bibitem{Wang2011}
\bibinfo{author}{Wang, L.~G.}, \bibinfo{author}{van~de Walle, A.} \&
  \bibinfo{author}{Alf{\`e}, D.}
\newblock \bibinfo{title}{Melting temperature of tungsten from twoab
  initioapproaches}.
\newblock \emph{\bibinfo{journal}{Phys. Rev. B Condens. Matter Mater. Phys.}}
  \textbf{\bibinfo{volume}{84}} (\bibinfo{year}{2011}).

\bibitem{lqb2021q}
\bibinfo{author}{Li, J.}, \bibinfo{author}{Qu, C.} \& \bibinfo{author}{Bowman,
  J.~M.}
\newblock \bibinfo{title}{Diffusion {M}onte {C}arlo with fictitious masses
  finds holes in potential energy surfaces} \textbf{\bibinfo{volume}{119}},
  \bibinfo{pages}{e1976426} (\bibinfo{year}{2021}).

\bibitem{Wood2017}
\bibinfo{author}{Wood, M.~A.} \& \bibinfo{author}{Thompson, A.~P.}
\newblock \bibinfo{title}{Quantum-accurate molecular dynamics potential for
  tungsten}.
\newblock \emph{\bibinfo{journal}{arXiv}}  (\bibinfo{year}{2017}).
\newblock \eprint{1702.07042}.

\bibitem{Perdew1996}
\bibinfo{author}{Perdew, J.~P.}, \bibinfo{author}{Burke, K.} \&
  \bibinfo{author}{Ernzerhof, M.}
\newblock \bibinfo{title}{Generalized gradient approximation made simple}.
\newblock \emph{\bibinfo{journal}{Phys. Rev. Lett.}}
  \textbf{\bibinfo{volume}{77}}, \bibinfo{pages}{3865--3868}
  (\bibinfo{year}{1996}).

\bibitem{Virtanen2020}
\bibinfo{author}{Virtanen, P.} \emph{et~al.}
\newblock \bibinfo{title}{{SciPy} 1.0: {Fundamental} algorithms for scientific
  computing in {P}ython}.
\newblock \emph{\bibinfo{journal}{Nat. Methods}} \textbf{\bibinfo{volume}{17}},
  \bibinfo{pages}{261--272} (\bibinfo{year}{2020}).

\bibitem{PingOng2019}
\bibinfo{author}{Ong, S.~P.}
\newblock \bibinfo{title}{Accelerating materials science with high-throughput
  computations and machine learning}.
\newblock \emph{\bibinfo{journal}{Comput. Mater. Sci.}}
  \textbf{\bibinfo{volume}{161}}, \bibinfo{pages}{143--150}
  (\bibinfo{year}{2019}).

\bibitem{Bartok2013}
\bibinfo{author}{Bart{\'{o}}k, A.~P.}, \bibinfo{author}{Kondor, R.} \&
  \bibinfo{author}{Cs{\'{a}}nyi, G.}
\newblock \bibinfo{title}{{On representing chemical environments}}.
\newblock \emph{\bibinfo{journal}{Phys. Rev. B}} \textbf{\bibinfo{volume}{87}},
  \bibinfo{pages}{184115} (\bibinfo{year}{2013}).

\bibitem{Bartok2015}
\bibinfo{author}{Bart{\'o}k, A.~P.} \& \bibinfo{author}{Cs{\'a}nyi, G.}
\newblock \bibinfo{title}{{G}aussian approximation potentials: A brief tutorial
  introduction}.
\newblock \emph{\bibinfo{journal}{Int. J. Quant. Chem.}}
  \textbf{\bibinfo{volume}{116}}, \bibinfo{pages}{1051--1057}
  (\bibinfo{year}{2015}).

\bibitem{Becker2013}
\bibinfo{author}{Becker, C.~A.}, \bibinfo{author}{Tavazza, F.},
  \bibinfo{author}{Trautt, Z.~T.} \& \bibinfo{author}{{Buarque De Macedo},
  R.~A.}
\newblock \bibinfo{title}{{Considerations for choosing and using force fields
  and interatomic potentials in materials science and engineering}}.
\newblock \emph{\bibinfo{journal}{Curr. Opin. Solid State Mater. Sci}}
  \textbf{\bibinfo{volume}{17}}, \bibinfo{pages}{277--283}
  (\bibinfo{year}{2013}).

\bibitem{Hale2018}
\bibinfo{author}{Hale, L.~M.}, \bibinfo{author}{Trautt, Z.~T.} \&
  \bibinfo{author}{Becker, C.~A.}
\newblock \bibinfo{title}{Evaluating variability with atomistic simulations:
  {The} effect of potential and calculation methodology on the modeling of
  lattice and elastic constants}.
\newblock \emph{\bibinfo{journal}{Model. Simul. Mater. Sci. Eng.}}
  \textbf{\bibinfo{volume}{26}}, \bibinfo{pages}{055003}
  (\bibinfo{year}{2018}).

\bibitem{Novikov2021}
\bibinfo{author}{Novikov, I.~S.}, \bibinfo{author}{Gubaev, K.},
  \bibinfo{author}{Podryabinkin, E.~V.} \& \bibinfo{author}{Shapeev, A.~V.}
\newblock \bibinfo{title}{The {MLIP} package: moment tensor potentials with
  {MPI} and active learning}.
\newblock \emph{\bibinfo{journal}{Machine Learning: Science and Technology}}
  \textbf{\bibinfo{volume}{2}}, \bibinfo{pages}{025002} (\bibinfo{year}{2021}).
\newblock \urlprefix\url{https://doi.org/10.1088/2632-2153/abc9fe}.

\bibitem{Jochym2018}
\bibinfo{author}{Jochym, P.~T.} \& \bibinfo{author}{Badger, C.}
\newblock \bibinfo{title}{jochym/elastic: Maintenance release}
  (\bibinfo{year}{2018}).
\newblock \urlprefix\url{https://doi.org/10.5281/zenodo.1254570}.

\bibitem{Togo2015}
\bibinfo{author}{Togo, A.} \& \bibinfo{author}{Tanaka, I.}
\newblock \bibinfo{title}{First principles phonon calculations in materials
  science}.
\newblock \emph{\bibinfo{journal}{Scripta Mater.}}
  \textbf{\bibinfo{volume}{108}}, \bibinfo{pages}{1--5} (\bibinfo{year}{2015}).

\bibitem{data_w}
\bibinfo{author}{Csanyi, G.}
\newblock \bibinfo{title}{Gaussian approximation potential for tungsten}
  (\bibinfo{year}{2022}).
\newblock \urlprefix\url{https://www.repository.cam.ac.uk/handle/1810/341742}.

\bibitem{data_si}
\bibinfo{author}{Csanyi, G.}
\newblock \bibinfo{title}{Research data: Machine learning a general-purpose
  interatomic potential for silicon} (\bibinfo{year}{2021}).
\newblock \urlprefix\url{https://www.repository.cam.ac.uk/handle/1810/317974}.

\end{thebibliography}
\vfill

\end{document}


\title{Ultra-fast interpretable machine-learning potentials}
\author{Stephen R. Xie}
\affiliation{Department of Materials Science and Engineering, University of Florida}
\affiliation{Quantum Theory Project, University of Florida}
\author{Matthias Rupp}
\affiliation{Department of Computer and Information Science, University of Konstanz, Germany}
\author{Richard G. Hennig}
\affiliation{Department of Materials Science and Engineering, University of Florida}
\affiliation{Quantum Theory Project, University of Florida}
\date{incomplete draft version of \today}

\keywords{machine learning, empirical potentials, force fields}

\maketitle{}



\section {Property data}

\begin{table}[h]
\caption{\emph{Derived properties.}}
\begin{tabular*}{\textwidth}{c @{\extracolsep{\fill}} rrrrrrrrrrrrr}
\hline
\multicolumn{1}{r}{\textbf{}} & \begin{tabular}[c]{@{}r@{}}Energy\\ (meV/atom)\end{tabular} & \begin{tabular}[c]{@{}r@{}}Forces\\ (eV/\AA)\end{tabular} & \begin{tabular}[c]{@{}r@{}}Phonons\\ (THz)\end{tabular} & \begin{tabular}[c]{@{}r@{}}$a_0$\\ (\AA)\end{tabular} & \begin{tabular}[c]{@{}r@{}}$C_{11}$\\ (GPa)\end{tabular} & \begin{tabular}[c]{@{}r@{}}$C_{12}$\\ (GPa)\end{tabular} & \begin{tabular}[c]{@{}r@{}}$C_{44}$\\ (GPa)\end{tabular} & \begin{tabular}[c]{@{}r@{}}$B$\\ (GPa)\end{tabular} & \begin{tabular}[c]{@{}r@{}}$E_{100}$\\ (eV)\end{tabular} & \begin{tabular}[c]{@{}r@{}}$E_{110}$\\ (eV)\end{tabular} & \begin{tabular}[c]{@{}r@{}}$E_{111}$\\ (eV)\end{tabular} & \begin{tabular}[c]{@{}r@{}}$E_V$ \\ (eV)\end{tabular} \\ \hline
\textbf{DFT} & \textbf{[225.4]} & \textbf{[1.496]} & \textbf{[2.254]} & \textbf{3.180} & \textbf{517.0} & \textbf{198.0} & \textbf{142.0} & \textbf{305.0} & \textbf{0.251} & \textbf{0.204} & \textbf{0.222} & \textbf{3.270} \\
$UF_{2}$  & 26.6   & 0.387  & 0.230   & 3.169 & 538.6    & 188.9    & 185.0    & 300.5 & 0.173     & 0.161     & 0.190     & 4.324 \\
LJ        & 110.0  & 1.400  & 3.914   & 3.105 & 506.2    & 600.9    & 599.4    & 566.7 & 0.345     & 0.325     & 0.348     & 4.334 \\
Morse     & 40.0   & 0.480  & 1.139   & 3.230 & 135.8    & 126.3    & 126.1    & 129.1 & 0.171     & 0.170     & 0.174     & 2.894 \\
$UF_{23}$ & 5.1    & 0.152  & 0.263   & 3.176 & 558.4    & 231.4    & 158.7    & 333.7 & 0.240     & 0.203     & 0.223     & 3.283 \\
EAM4      & 88.0   & 0.803  & 0.301   & 3.143 & 525.3    & 206.5    & 163.8    & 311.0 & 0.184     & 0.159     & 0.224     & 3.816 \\
MTP       & 16.5   & 0.146  & 0.376   & 3.168 & 616.2    & 274.5    & 164.1    & 388.4 & 0.236     & 0.192     & 0.259     & 3.232 \\
SNAP      & 14.2   & 0.189  & 0.270   & 3.166 & 653.8    & 335.4    & 124.2    & 433.4 & 0.227     & 0.196     & 0.261     & 2.048 \\
qSNAP     & 9.9    & 0.167  & 0.256   & 3.176 & 497.0    & 179.2    & 101.9    & 281.1 & 0.249     & 0.202     & 0.245     & 2.574 \\
GAP       & 6.2    & 0.169  & 0.291   & 3.178 & 596.4    & 253.7    & 142.0    & 363.5 & 0.268     & 0.216     & 0.177     & 3.342 \\
\hline
\end{tabular*}
\end{table}

\section {Model parameters}

\begin{table}[H]
\caption{\emph{UF Potential hyperparameters selected in this work.}}
\label{tab:my-table}
\begin{tabular*}{\textwidth}{l @{\extracolsep{\fill}} rrrrcrrrrrcr}
\hline
 & \multicolumn{6}{l}{two-body} & \multicolumn{6}{l}{three-body} \\ \hline
 & \multicolumn{1}{r}{\begin{tabular}[c]{@{}r@{}}$r_{\text{min},2}$\\ (\AA)\end{tabular}} & \multicolumn{1}{r}{\begin{tabular}[c]{@{}r@{}}$r_{\text{cut},2}$\\ (\AA)\end{tabular}} & \begin{tabular}[c]{@{}r@{}}knot\\ spacing\end{tabular} & \begin{tabular}[c]{@{}r@{}}basis \\ functions\end{tabular} & symmetry & $\lambda_2$ & \begin{tabular}[c]{@{}r@{}}$r_{\text{min},3}$\\ (\AA)\end{tabular} & \begin{tabular}[c]{@{}r@{}}$r_{\text{cut},3}$\\ (\AA)\end{tabular} & \begin{tabular}[c]{@{}r@{}}knot\\ spacing\end{tabular} & \begin{tabular}[c]{@{}r@{}}basis \\ functions\end{tabular} & symmetry & $\lambda_3$ \\  \hline
$\text{UF}_2$ & 1.5 & 5.5 & linear & 25 & i-j = j-i & 1E-08 & - & - & - & - & - & - \\
$\text{UF}_{2,3}$ & 1.5 & 5.5 & linear & 25 & i-j = j-i & 1E-08 & \multicolumn{1}{r}{1.5} & \multicolumn{1}{r}{4.25} & linear & \multicolumn{1}{r}{915} & i-j-k = i-k-j & \multicolumn{1}{r}{1E-08} \\  \hline
\end{tabular*}
\end{table}

\begin{table}[H]
\caption{\emph{SNAP/qSNAP hyperparameters selected in this work.}}
\label{tab:my-table}
\begin{tabular*}{\textwidth}{l @{\extracolsep{\fill}} rrrrrr}
\hline
 & \begin{tabular}[c]{@{}r@{}}rcutfac\\ (\AA)\end{tabular} & twojmax & rfac0 & rmin0 & quadraticflag & bzeroflag \\ \hline
SNAP & 5.5 & 8 & 0.99363 & 0 & 0 & 0 \\
qSNAP & 5.5 & 6 & 0.99363 & 0 & 1 & 0 \\ \hline
\end{tabular*}
\end{table}

\begin{table}[H]
\caption{\emph{GAP hyperparameters selected in this work.}}
\label{tab:my-table}
\begin{tabular*}{\textwidth}{l @{\extracolsep{\fill}} rrrrrrrrrrr}
\hline
 & \begin{tabular}[c]{@{}r@{}}cutoff\\ (\AA)\end{tabular} & l\_max & n\_max & atom\_sigma & zeta & cutoff\_transition\_width & delta & f0 & n\_sparse & covariance\_type & sparse\_method \\ \hline
GAP & 5.5 & 8 & 8 & 0.5 & 4 & 0.5 & 1 & 0 & 200 & dot\_product & cur\_points \\ \hline
\end{tabular*}
\end{table}

\pagebreak

\section {Hyperparameter exploration in UF potentials}

\begin{table}[h]
\caption{\emph{Basis functions and error vs. cutoff radius in $\text{UF}_{2,3}$ potential.}}
\begin{tabular*}{\textwidth}{l @{\extracolsep{\fill}} lrrrrrr} \hline
\multicolumn{1}{l}{$r_{\text{cut}}$} & $\lambda$ & \multicolumn{1}{r}{\begin{tabular}[c]{@{}r@{}}RMSE$_{\mathrm{E}}$\\ (meV/atom)\end{tabular}} & \multicolumn{1}{r}{\begin{tabular}[c]{@{}r@{}}RMSE$_{\mathrm{F}}$\\ (meV/\AA)\end{tabular}} & \multicolumn{1}{r}{$\|c \neq 0 \|$} & \multicolumn{1}{r}{$\| c \|$} & \multicolumn{1}{r}{\begin{tabular}[c]{@{}r@{}}symmetry\\ mask\end{tabular}} & \multicolumn{1}{r}{\begin{tabular}[c]{@{}r@{}}cutoff\\ mask\end{tabular}} \\ \hline
4.00 & $10^{-6}$ & 7.380 & 0.205 & 924 & 2000 & 353 & 723 \\
4.00 & $10^{-7}$ & 7.311 & 0.203 & 924 & 2000 & 353 & 723 \\
4.00 & $10^{-8}$ & 7.309 & 0.203 & 924 & 2000 & 353 & 723 \\
4.25 & $10^{-6}$ & 6.335 & 0.179 & 915 & 2000 & 368 & 717 \\
4.25 & $10^{-7}$ & 6.344 & 0.176 & 915 & 2000 & 368 & 717 \\
4.25$^*$ & $10^{-8\,*}$ & 6.340 & 0.176 & 915 & 2000 & 368 & 717 \\
4.50 & $10^{-6}$ & 42.973 & 1.553 & 905 & 2000 & 381 & 714 \\
4.50 & $10^{-7}$ & 6.116 & 0.171 & 905 & 2000 & 381 & 714 \\
4.50 & $10^{-8}$ & 6.115 & 0.171 & 905 & 2000 & 381 & 714 \\
4.75 & $10^{-6}$ & 5.731 & 0.168 & 903 & 2000 & 386 & 711 \\
4.75 & $10^{-7}$ & 5.576 & 0.163 & 903 & 2000 & 386 & 711 \\
4.75 & $10^{-8}$ & 5.575 & 0.163 & 903 & 2000 & 386 & 711 \\
5.00 & $10^{-6}$ & 5.326 & 0.159 & 899 & 2000 & 393 & 708 \\
5.00 & $10^{-7}$ & 5.190 & 0.156 & 899 & 2000 & 393 & 708 \\
5.00 & $10^{-8}$ & 5.189 & 0.156 & 899 & 2000 & 393 & 708 \\
5.25 & $10^{-6}$ & 35.824 & 1.414 & 889 & 2000 & 406 & 705 \\
5.25 & $10^{-7}$ & 5.208 & 0.154 & 889 & 2000 & 406 & 705 \\
5.25 & $10^{-8}$ & 5.210 & 0.154 & 889 & 2000 & 406 & 705 \\
5.50 & $10^{-6}$ & 32.388 & 1.310 & 888 & 2000 & 410 & 702 \\
5.50 & $10^{-7}$ & 5.106 & 0.152 & 888 & 2000 & 410 & 702 \\
5.50 & $10^{-8}$ & 5.107 & 0.152 & 888 & 2000 & 410 & 702 \\ \hline
\end{tabular*}
\end{table}

\begin{figure}[H]
\includegraphics[width=6.5in]{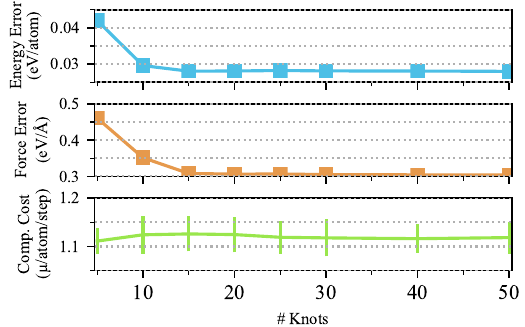}
\caption{\emph{Convergence in two-body interaction properties with number of knots. Energy and force errors quickly converge with the number of knots, which, in turn, determine the number of basis functions. The computational cost of evaluation does not scale wtih the number of knots due to compact support.}}
\end{figure}

\section{Reproduction of pair and two-and-three-body potentials with UF potentials}

\begin{figure}[H]
\includegraphics[width=\columnwidth]{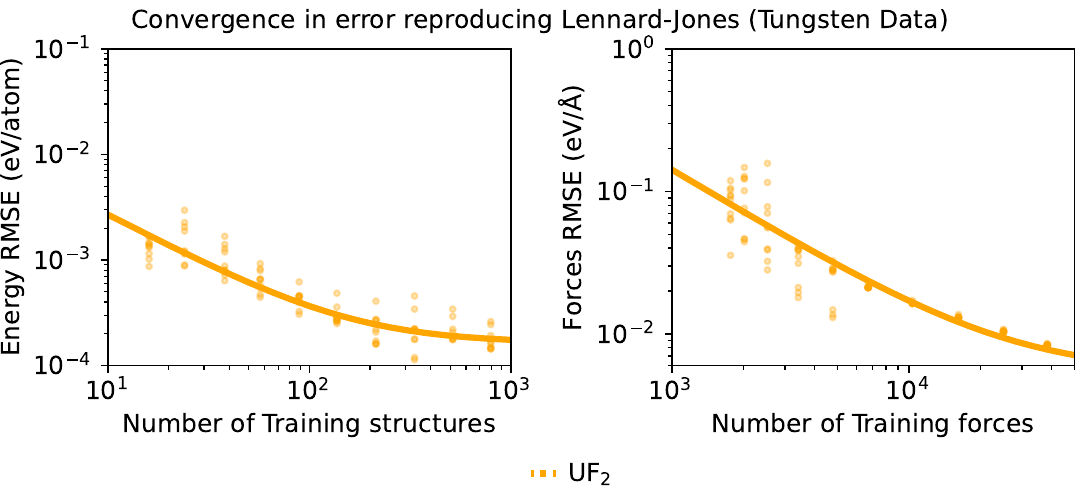}
\caption{\emph{Convergence in errors in reproducing the Lennard-Jones potential with the two-body UF potential. The dataset used to construct these learning curves contain various elemental tungsten configurations (e.g. bcc, vacancy, gamma surfaces) and was previously constructed by Szlachta et al. to fit GAP potentials~\cite{Szlachta2014}.}}
\end{figure}

\begin{figure}[H]
\includegraphics[width=\columnwidth]{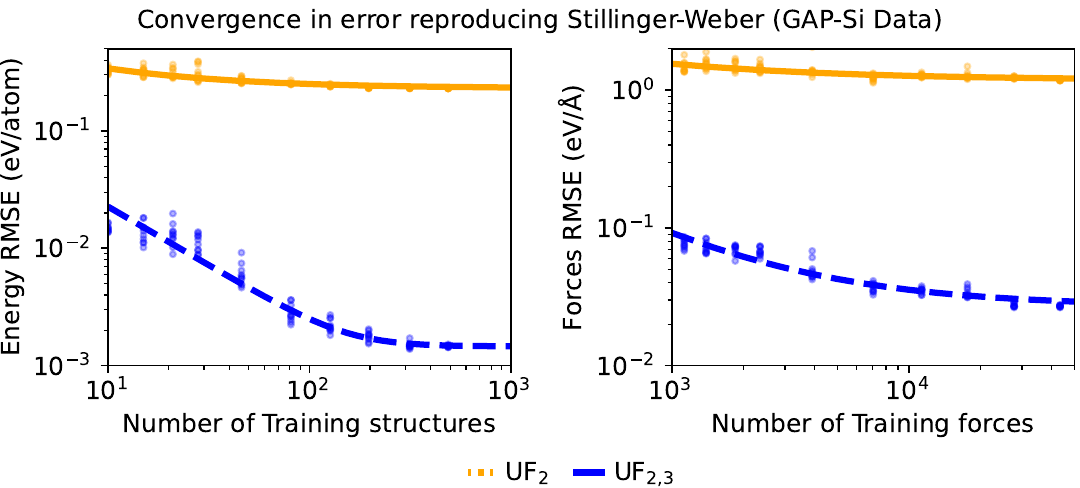}
\caption{\emph{Convergence in errors in reproducing the Stillinger-Weber potential with the $\text{UF}_2$ and $\text{UF}_{2,3}$ potentials. The dataset used to construct these learning curves contain various elemental silicon configurations (e.g. diamond, hexagonal, $\beta$-Sn, amorphous) and was previously constructed by Bart{\'{o}}k et al. to fit GAP potentials~\cite{Bartok2018}.}}
\td{angular-dependent terms cannot be reproduced by 2-body potential. as a test.}}
\end{figure}

\section{Additional Figures}

\begin{figure}[H]
\includegraphics[width=\columnwidth]{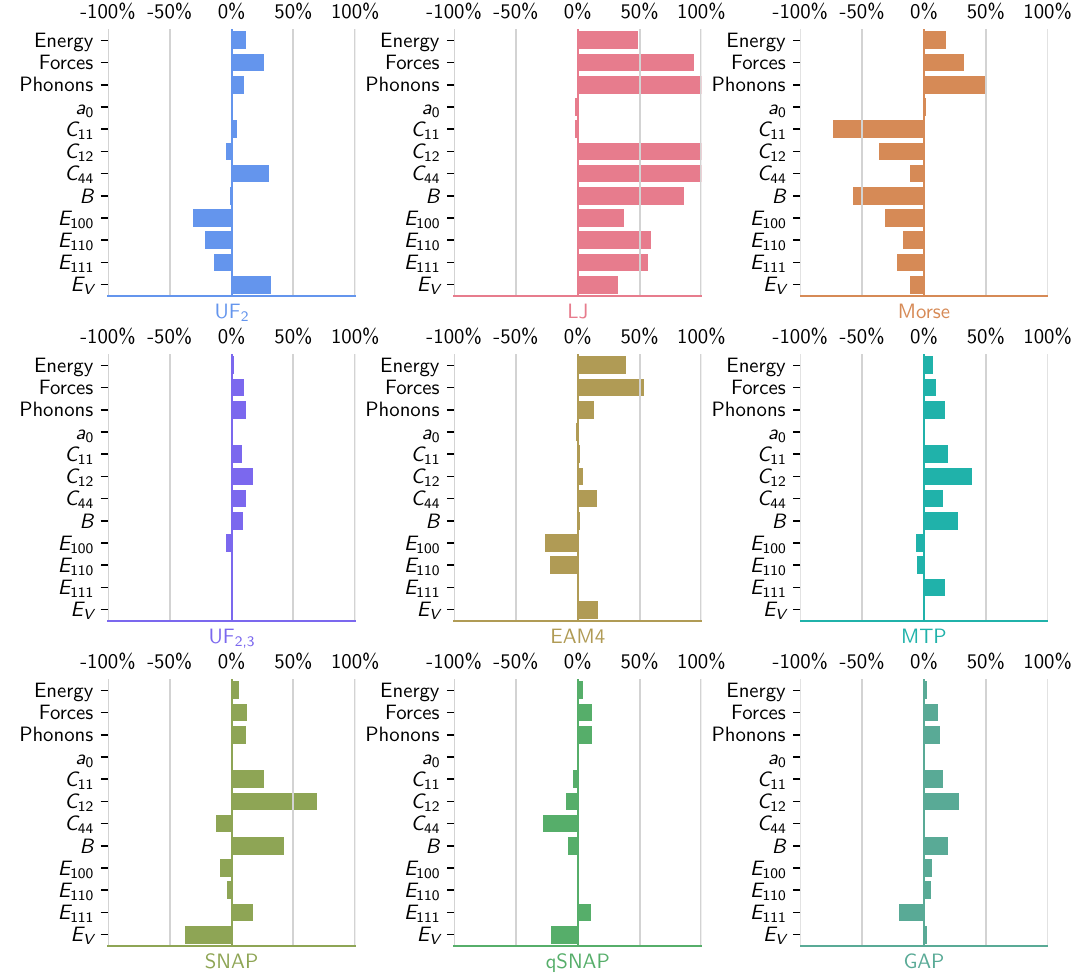}
\caption{
\emph{Performance for derived quantities} of seven potentials relative to the DFT reference for bcc tungsten. Energy, force, and phonon spectra error are percent RMSE normalized by the sample standard deviation of the reference values. Other errors are percentage errors. The $\text{UF}_2$ potential achieves an accuracy approaching that of SNAP and qSNAP, while the $\text{UF}_{2,3}$ potential achieves an accuracy comparable to MTP and GAP.
}
\end{figure}

\begin{figure}[t]
\centering
\includegraphics[width=3.5in]{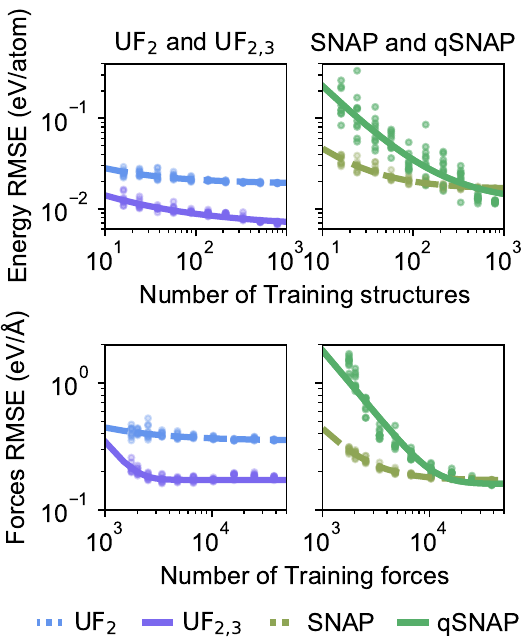}
\caption{
\emph{Energy and force learning curves for UF and SNAP potentials fit to bcc tungsten.} 
Shown are out-of-sample prediction errors (dots) for energies (top) and forces (bottom) for training sets of increasing size, with 5 repetitions per size.
Fitted curves (solid lines) are soft-plus functions that capture both the initial linear slope in log-log space and the observed saturation.
Simpler potentials saturate earlier, but with higher error than more complex potentials (UF$_2$\ vs.\ SNAP; UF$_2$\ vs.\ UF$_3$; SNAP\ vs.\ QSNAP), outperforming them when training data is limited.
}
\pagebreak

\label{fig:learning}
\end{figure}

\pagebreak

\bibliography{uf_references}
\vfill